\documentclass[fleqn,usenatbib]{mnras}

\usepackage[T1]{fontenc}

\usepackage{multirow}
\usepackage{tablefootnote}
\DeclareRobustCommand{\VAN}[3]{#2}
\let\VANthebibliography\thebibliography
\def\thebibliography{\DeclareRobustCommand{\VAN}[3]{##3}\VANthebibliography}

\usepackage{multicol}
\usepackage{graphicx}	
\usepackage{amsmath}	
\usepackage{amssymb}	

\usepackage{amsmath}

\usepackage{newtxtext,newtxmath}

\title[Unraveling the evolution of hot Jupiter systems]{Unraveling the evolution of hot Jupiter systems under the effect of tidal and magnetic interactions and mass loss}

\author[Y. A. Lazovik]{
Yaroslav A. Lazovik$^{1,2}$\thanks{E-mail: yaroslav.lazovik@gmail.com}
\\
$^{1}$Lomonosov Moscow State University, Faculty of Physics, 1 Leninskie Gory, bldg.2, Moscow, 119991, Russia\\
$^{2}$Sternberg Astronomical Institute, Lomonosov Moscow State University, Universitetsky pr. 13, Moscow, 119234, Russia
}
\date{Accepted XXX. Received YYY; in original form ZZZ}

\pubyear{2022}

\begin{document}
\label{firstpage}
\pagerange{\pageref{firstpage}--\pageref{lastpage}}
\maketitle

\begin{abstract}
Various interactions affect the population of close-in planets. Among them, the tidal and magnetic interactions drive orbital decay and star-planet angular momentum exchange, leading to stellar spin-up. As a result of the above processes, a planet may initiate the mass transfer to the host star once it encounters the Roche limit. Another mechanism providing substantial mass loss is associated with the atmospheric escape caused by photoevaporation followed by orbital expansion, which is thought to be important for hot Neptunes and super-Earths. Thus, the fraction of the initial number of hot Jupiters may transform into lower-mass planets through the Roche-lobe overflow (RLO) phase and continue secular evolution under the effect of photoevaporation. In the present paper, we compile the latest prescriptions for tidal and magnetic migration and mass-loss rates to explore the dynamics of hot Jupiter systems. We study how the implemented interactions shape the orbital architecture of Jovian planets and whether their impact is enough to reproduce the observational sample. Our models suggest that the tidal interaction is able to generate the upper boundary of the hot Jupiter population in the mass--separation diagram. To recreate the sub-Jovian desert, we need to make additional assumptions regarding the RLO phase or the influence of the protoplanetary disc's inner edge on the initial planetary location. According to our estimates, 12--15 \% of hot Jupiters around solar-mass stars have been engulfed or become lower-mass planets. 0.20--0.25 \% of the present-day giant planet population undergoes decay intense enough to be detected with modern facilities.

\end{abstract}




\section{Introduction}

Since the discovery of 51 Peg b (\citealt{Mayor}), exploring the evolution of hot Jupiters remains one of the leading directions in the field of exoplanet studies. High detection efficiency provides the rapid growth of the observational sample, allowing testing theories of planetary formation. In addition, various interactions are likely to play a significant role in the orbital architecture of Jovian planets throughout their lifetime, which is why the recent implementations of the population synthesis approach (\citealt{Emsenhuber1, Emsenhuber2,Schlecker, Burn}) follow the long-term evolution to reproduce the patterns of planetary statistics.

Among different processes in planetary systems, tidal force is believed to be one of the most substantial in its impact of the dynamics of hot Jupiters. The pioneering theory of tidal interaction has been developed in \cite{Darwin}. In the aforementioned study, the tidal adjustment has been considered in the form of two symmetric bulges in hydrostatic equilibrium. Owing to friction, the bulges are shifted with respect to the line joining the center of a star and its companion, which produces tidal torque. Furthermore, the corresponding constant time lag model has been investigated and refined in \cite{Alexander}, \cite{Hut}, and \cite{Eggleton}. The prominent improvement has been made in \cite{Zahn1,Zahn2,Zahn3}, who considered the wavelike contribution to the tidal response. Subsequent decades of research have resulted in major progress in understanding the mechanisms underlying tidal excitation and dissipation. Based on the observational data, \cite{Hansen1, Hansen2} and \cite{Penev} constrained equilibrium tide dissipation, which in the case of fluid bodies is thought to arise due to the convective motions acting as an effective viscosity (for rocky planets, different mechanisms provide tidal dissipation, see \cite{Henning}). However, the latest hydrodynamical simulations have shown that, due to the interaction between tidal flows and convection, the turbulent viscosity is substantially reduced (\citealt{Duguid1,Duguid2,VidalBarker1,VidalBarker}), resulting in low migration rates unable to explain many observed features, in particular, the migration rates of WASP-12 (\citealt{Maciejewski,Yee,Turner}). One way to overcome this discrepancy is to extend the tidal formalism by considering the additional effects responsible for tidal dissipation, such as magnetic diffusivity (\citealt{Wei}). Alternatively, the importance of equilibrium tide in the context of planetary orbital evolution may turn out to be negligible compared to the dynamical tides. Thus, the dominant mechanism for rapid migration may be associated with the dissipation of gravity waves in the stellar radiative region. In this regard, several regimes of gravity wave damping are distinguished, namely linear (\citealt{Goodman}), weakly nonlinear (\citealt{Kumar,BarkerOgilvie1,Weinberg,Essick,ICB}), and strongly nonlinear (\citealt{Goodman,Ogilvie00,BarkerOgilvie, Ipch, Barker1, Barker}) regimes. Inertial waves, which propagate in the convective zone, affect the dynamics of massive planets around the young, rapidly rotating stars. Applying the frequency-averaged formalism by \cite{Ogilvie1}, \cite{Mathis} calculated inertial wave dissipation rates for a simplified homogeneous two-layer stellar model with a radiative core and a convective envelope. \cite{Barker2022} made another step forward by estimating the binary circularization periods following the prescriptions for a more complex heterogeneous model reported in \cite{Barker}.

In the previous paper, we focused on the evolution of hot Jupiters under the effect of tidal dissipation (\citealt{Lazovik}, hereafter L21). To do so, we adopted the tidal formalism from \cite{Barker}, allowing us to study the dissipation of equilibrium tide, inertial waves, and gravity waves over the wide parameter space. Based on the obtained results, we simulated the hot Jupiter population and derived the statistics of planetary infalls within the Galactic thin disc. According to our estimates, 11 -- 21\% of the initial number of hot Jupiters (hereafter, we define hot Jupiters as the planets with the orbital period $P_\mathrm{orb}$ below 10 days, and the planetary mass $M_\mathrm{pl}$ between 0.3 and 10 $M_\mathrm{J}$, where the subscript 'J' denotes the Jovian units) around the solar-type stars undergo engulfment before the host's main sequence (MS) termination.

Nevertheless, tidal dissipation is not the only mechanism affecting planetary migration. Relative motion between a planet and magnetized ambient stellar wind causes magnetic interaction. If the planetary magnetosphere is sustained, the dipolar regime develops (\citealt{Strugarek1,Strugarek2,Strugarek3,Strugarek4}). The extensive study by \cite{Ahuir} has shown that the dipolar regime may dominate the secular evolution of low-mass planets orbiting slow rotators. The situation is less clear when the orbital separation is small enough for the planetary intrinsic magnetic field to be suppressed. If the magnetic configuration is stable and interaction proceeds without the breakdown of the circuit, the magnetic torque accelerates the migration of a planet. The corresponding regime is called unipolar (\citealt{Laine2}). However, as discussed in \cite{Lai}, the flux tube is likely to break up when the azimuthal twist is too large, disconnecting the linkage between the components, which is why the existence of unipolar interaction is still under debate.

In addition, close-in planets are exposed to strong stellar irradiation, which drives photoevaporation and associated outward migration shaping the orbital architecture of the exoplanet population (\citealt{Boue, Rao1}). While being effective for super-Earths (\citealt{Fujita}) and sub-Neptunes (\citealt{OwenJackson,OwenWu,Rogers}), photoevaporation may not significantly alter the masses of Jovian planets (\citealt{Murray-Clay,OwenWu}), although the hydrodynamic simulations have shown that the close-in gas giants of less than a Jupiter mass are likely to be affected by the thermally-driven outflow (\citealt{Salz,Caldiroli}). Besides, hot Jupiters lose mass during Roche-lobe overflow (RLO) leading to a transformation into a lower-mass planet. This so-called stable accretion scenario has been investigated in \cite{Valsecchi,Valsecchi1,Jackson}.

Finally, it is worth mentioning the high-eccentricity processes, such as secular chaos, planet-planet scattering, and planet-planet Kozai migration (\citealt{FordRasio,Wu,Naoz,Beauge,ValsecchiRasio,ValsecchiRasio1,Attia,Wang}), that may explain the proximity of hot Jupiters without resorting to a disc migration scenario. As demonstrated in \cite{Gu}, tidal dissipation within hot Jupiter in a modestly eccentric orbit and the associated internal heating can drive atmospheric expansion, resulting in the early onset of RLO. Although the aforementioned mechanisms are beyond the scope of the present work, they may play a significant role in the formation and dynamics of close-in giant planets.

We extend the model of planetary migration developed in L21 by taking into account magnetic interaction, photoevaporation, and RLO. The main goal of our research is to examine the importance of various properties in the orbital evolution of hot Jupiters around solar-type stars. We trace the transformation of the synthetic planetary population and study how the choice of the Roche limit parametrization may alter the outcome. This paper is structured as follows. In Sec.~\ref{sec:model}, we introduce our approach and analyze in detail the interactions considered in the present work. The impact of different properties of a star-planet system (e.g., stellar and planetary mass, initial semi-major axes and stellar rotation, planetary magnetic field strength) is characterized in Sec.~\ref{sec:outline}. In Sec.~\ref{sec:popsynth}, we study the evolution of the hot Jupiter population within the framework of our model. The results are discussed and summarized in Sec.~\ref{sec:summary}.

\section{Model description}
\label{sec:model}
In the present section, we briefly mention the prescriptions that remained unchanged with respect to L21 and then focus on the new features of our model. We consider star-planet systems composed of a spherically-symmetric uniformly rotating solar-mass star and a point-mass planet in a circular equatorial orbit. Unlike L21, we do not concentrate on simulating the orbital evolution until the main-sequence (MS) termination and extend our simulation to the subgiant phase until either the planet merges with the host or loses most of its gaseous envelope and its mass fraction drops below the critical value of 1\%. We assume the planetary core mass $M_\mathrm{c} = 10\;M_{\oplus}$. Note that \cite{Valsecchi1} and \cite{Jackson} reported that, owing to the impact of $M_\mathrm{c}$ to the mass--radius relation, the core mass might play a key role in the process and the outcome of stable accretion. Thus, hot Jupiters with massive cores completely shed their envelope during RLO. On the contrary, the planets with lower-mass cores detach from the Roche limit, $a_\mathrm{R}$, holding a small fraction of the envelope. By adopting $M_\mathrm{c} = 10\;M_{\oplus}$, we follow the core accretion hypothesis by \cite{Pollack}. We recall that, according to \cite{Batygin}, smaller cores cannot induce gas accretion intense enough to produce hot Jupiters within the disc's lifetime, while the formation of higher mass cores requires the enhancement of the initial solid density of the protoplanetary disc (\citealt{Pollack}). Besides, the lack of ultra-short period rocky planets with massive core (which are expected to remain after the RLO termination, see \cite{Valsecchi1}) may potentially imply the relatively low contribution of hot Jupiters with $M_\mathrm{c} > 15\;M_{\oplus}$ to the overall population.

\subsection{Stellar model} \label{subsec:stellar}
\label{subsec:star}

As in L21, we compute stellar models using evolutionary code MESA (\citealt{MESA1,MESA2,MESA3,MESA4,MESA5}) and inlist files from \cite{Gossage}. The angular momentum of isolated stars decreases under the braking law proposed in \cite{Matt} and \cite{Amard}. Following \cite{Rebull}, the rotation rate is fixed during the disc's lifetime. One significant difference from L21 concerns the estimation of the disc dissipation timescale $\tau_\mathrm{disc}$. In the present study, we use calibration from \cite{Tu}:
\begin{equation}
    \tau_\mathrm{disc} = 13.5 \; \left(\frac{\Omega_{ *, \; \rm 0}}{ {\Omega_{\odot}}}\right)^{-0.5} \, {\rm Myr}, 
	\label{eq:disc}
\end{equation}
with $\Omega_{*, \;\rm 0}$ the initial stellar spin. The symbol $\odot$ denotes the solar value.

\subsection{Orbital evolution} \label{subsec:migration}
The planetary migration rate is specified as a sum of three components:
\begin{equation}
    \frac{\dot{a}}{a} = \left(\frac{\dot{a}}{a}\right)_\mathrm{t} + \left(\frac{\dot{a}}{a}\right)_\mathrm{m} + \left(\frac{\dot{a}}{a}\right)_\mathrm{ML}, 
	\label{eq:migr}
\end{equation}
where $a$ is the semi-major axis of a planet, $\left(\frac{\dot{a}}{a}\right)_\mathrm{t}$ represents the tidal contribution, $\left(\frac{\dot{a}}{a}\right)_\mathrm{m}$ --- the magnetic contribution, and  $\left(\frac{\dot{a}}{a}\right)_\mathrm{ML}$  --- the contribution arising due to mass loss. 

Tidal and magnetic interactions lead to a redistribution of angular momentum, affecting the rotation rate of a star $\Omega_{*}$. The evolution of stellar spin follows the expression:
\begin{equation}
    \dot{\Omega}_{*} = \frac{1}{I_{*}} \left(\Gamma_\mathrm{wind} - \Omega_{*} \dot{I}_{*} - \frac{1}{2} I_\mathrm{pl} n \left(\frac{\dot{a}}{a} \right)_\mathrm{tm} \right).   
	\label{eq:rot}
\end{equation}
 $\left(\frac{\dot{a}}{a} \right)_\mathrm{tm} = \left(\left(\frac{\dot{a}}{a}\right)_\mathrm{t} + \left(\frac{\dot{a}}{a}\right)_\mathrm{m} \right)$, $n$ is the orbital angular frequency, $\Gamma_\mathrm{wind}$ is the wind torque, $I_{*}$ is the stellar moment of inertia, $I_\mathrm{pl} = M_\mathrm{pl} a^2$,  $M_{*}$ is the stellar mass. Here, we neglect the variation of stellar rotation rate due to the mass exchange.
 
\begin{figure}
	\includegraphics[width=\columnwidth]{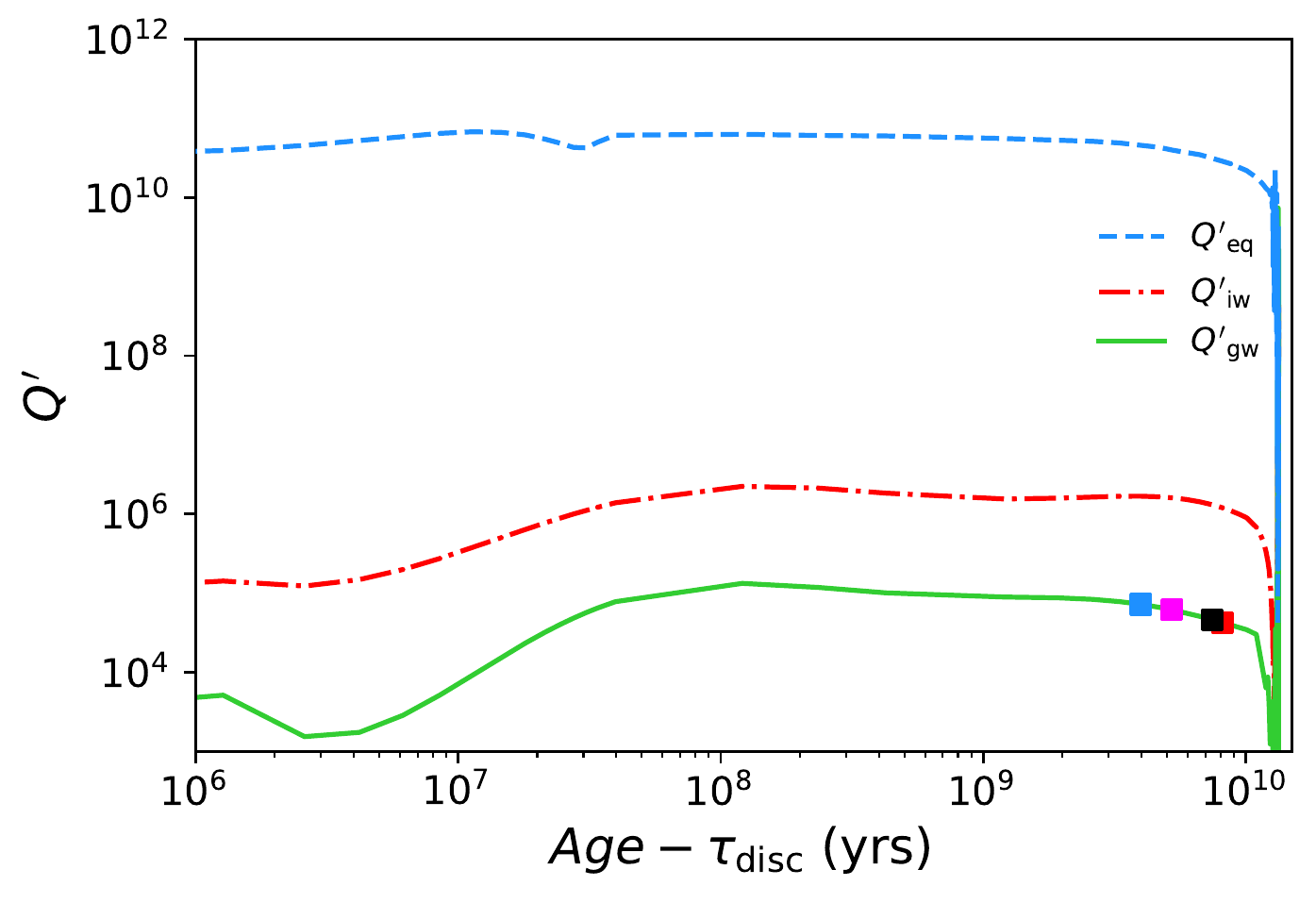}
    \caption{Tidal quality factor as a function of time after disc dissipation ($M_{*}$ = 1.0$ \; M_{\odot}$, [Fe/H] = +0.2 dex,  $P_\mathrm{orb}$ = 1 day, $P_\mathrm{rot}$ = 4.5 days). Stellar rotation and planetary orbital separation are fixed. Lines, from top to bottom, represent equilibrium tide, inertial waves, and gravity waves, respectively. Red, black, magenta, and blue squares (from left to right) correspond to the onset of gravity wave dissipation for a planet with $M_\mathrm{pl}$ = 0.3 $M_\mathrm{J}$, 1 $M_\mathrm{J}$, 3 $M_\mathrm{J}$, and 10 $M_\mathrm{J}$, respectively.}
    \label{fig0}
\end{figure}
 
\subsubsection{Tidal contribution}
Tidal dissipation leads to the redistribution of angular momentum in a star-planet system, resulting in planetary migration. Since the rotation of massive planets is typically synchronized within a short timescale (\citealt{Guillot}), we neglect the impact of planetary tides on secular evolution. Similar to L21, we consider three types of tides, namely equilibrium tide, inertial waves, and gravity waves. Each of the following types is characterized by the tidal quality factor $Q'$, which links the efficiency of tidal energy dissipation with the associated migration rate via the expression:
\begin{equation}
    \left(\frac{\dot{a}}{a}\right)_\mathrm{t} = \frac{\Omega_{*} - n}{|\Omega_{*} - n|}\frac{9n}{2} \left(\frac{M_\mathrm{pl}}{M_{*}} \right) \left(\frac{R_{*}}{a} \right)^5 \frac{1}{Q'}. 
	\label{eq:orbit7}
\end{equation}
 $R_{*}$ is the stellar radius. The tidal quality factors associated with each type of tidal response ($Q'_\mathrm{eq}$, $Q'_\mathrm{iw}$, and $Q'_\mathrm{gw}$ for equilibrium tide, inertial waves, and gravity waves, respectively) are derived according to \cite{Barker}. The resulting tidal quality factor is expressed by:

\begin{equation}
    \frac{1}{Q'} = \frac{1}{Q'_\mathrm{eq}} + \frac{1}{Q'_\mathrm{iw}} + \frac{1}{Q'_\mathrm{gw}}
    \label{tide}
\end{equation}
 
 We remind that inertial waves are being excited when the orbital period $P_\mathrm{orb}$ is over half the rotation period of a star $P_\mathrm{rot}$, while gravity wave damping occurs if the planetary mass is above the critical value  $M_\mathrm{crit}(M_{*},a,t)$, which is satisfied toward the end of the MS lifetime. For details, regarding the calculation of $Q'$, we refer to \cite{Lazovik} and \cite{Barker}. Fig.~\ref{fig0} compares the tidal quality factors due to each tidal mechanism as a function of time after disc dissipation in the case of a solar-mass star with $P_\mathrm{rot}$ = 4.5 days hosting a planet on a one-day orbit. One can see that, when enabled, the dissipation of dynamical waves (gravity and inertial waves) makes a dominant contribution as the corresponding tidal quality factors are in the range between [$10^3$, $10^7$], while the tidal quality factor representing equilibrium tide is above $10^{10}$ until TAMS. When a star evolves off the main sequence, the efficiency of equilibrium tide dissipation begins to vary over a wide range, and at some moment, it even surpasses the dynamical wave damping.  Another prominent feature is the dependence of the onset of gravity wave dissipation, illustrated by the squares, on the planetary mass. For the most massive planets considered here ($M_\mathrm{pl}$ = 10 $M_\mathrm{J}$), gravity wave breaking begins at t $\sim$ 3 Gyr, while for the planets with $M_\mathrm{pl}$ = 0.3 $M_\mathrm{J}$, it occurs shortly before TAMS. This indicates that more massive planets have more time for rapid migration, which is shown in subsection~\ref{subsec:axmass}.
 
\subsubsection{Magnetic contribution}

\begin{figure}
	\includegraphics[width=\columnwidth]{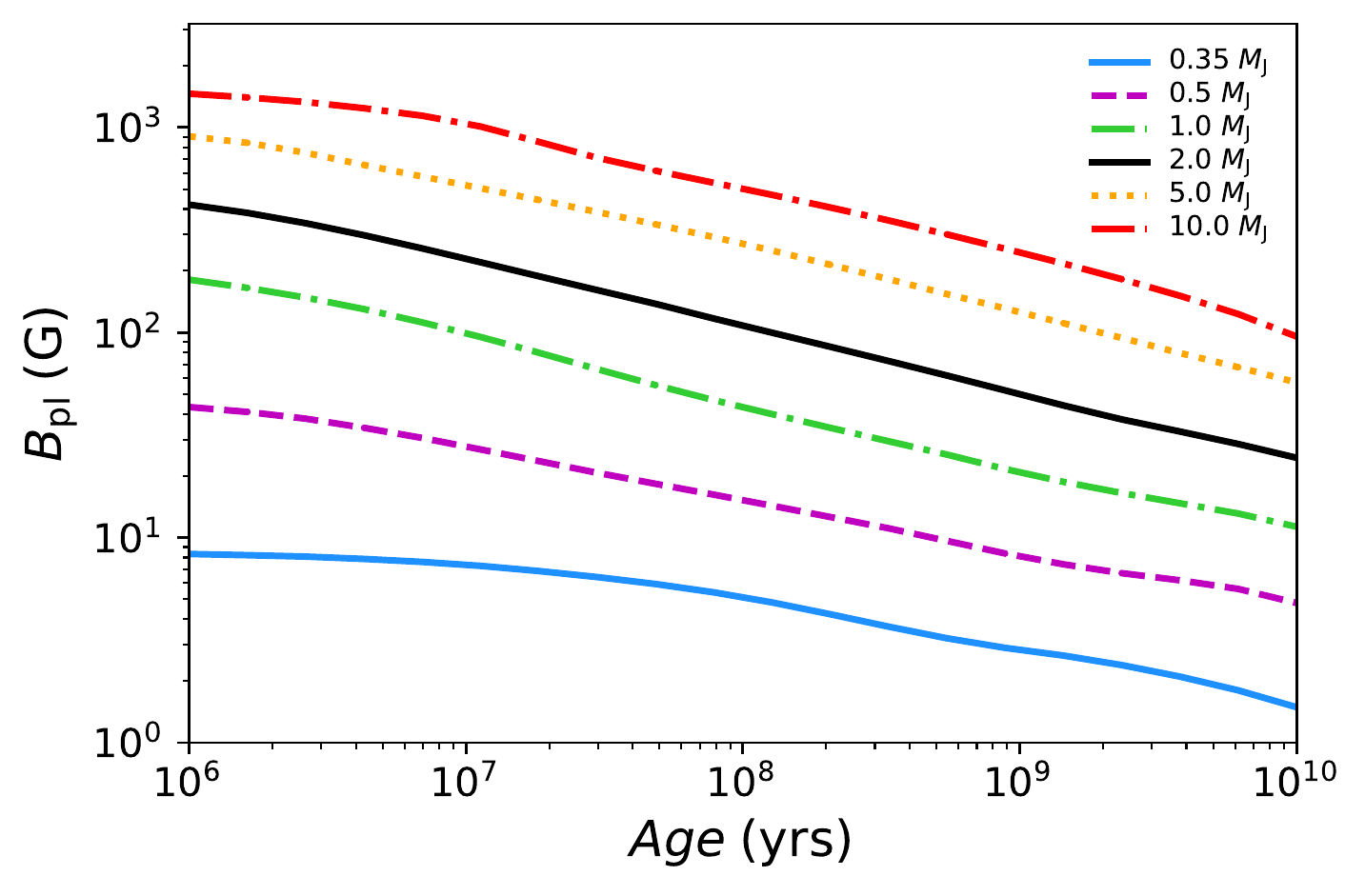}
    \caption{The equatorial magnetic field at the planetary surface obtained following~\protect\cite{Hori}.}
    \label{fig1}
\end{figure} 

The planetary orbital motion produces perturbations in the magnetized medium, which propagate in the form of magneto-hydrodynamic waves away from the planet, carrying its angular momentum and energy (\citealt{Strugarek2}). The intensity of this process depends on the local Alfven Mach number $M_\mathrm{a}$, the magnetic topology, and the effective area of the planetary obstacle $A_\mathrm{eff}$. In this work, the magnetic contribution is represented by the dipolar torque taken from \cite{Ahuir}:
\begin{equation}
     \left(\frac{\dot{a}}{a}\right)_\mathrm{m} = \frac{\Omega_{*} - n}{|\Omega_{*} - n|}\frac{2}{I_\mathrm{pl}n} A_\mathrm{eff} c_\mathrm{d} p_\mathrm{tot},
    \label{eq:mag}
\end{equation}
where $c_\mathrm{d} \approx \frac{M_\mathrm{a}}{\sqrt{M_\mathrm{a}^2+1}}$ is the drag coefficient, and $p_\mathrm{tot}$ is the total local pressure of the ambient wind. The effective obstacle area is:
\begin{equation}
     A_\mathrm{eff} = \begin{cases}
 \left(10.8 M_\mathrm{a}^{-0.56}  \Lambda_\mathrm{p}^{0.28} \right)\pi R_\mathrm{pl}^2, & \Lambda_\mathrm{p} > 1;\\
\pi R_\mathrm{pl}^2, & \Lambda_\mathrm{p} \leq 1;
 \end{cases}
    \label{eq:obst}
\end{equation}
with $\Lambda_\mathrm{p} = \frac{B_\mathrm{pl}^2}{2 \mu_{0} p_\mathrm{tot}}$. $B_\mathrm{pl}$ is the equatorial field at the planetary surface, $R_\mathrm{pl}$ is the planetary radius, and $\mu_{0}$ is the permeability of free space. Note that $\Lambda_\mathrm{p}$ is above unity when a planet sustains its magnetosphere, which significantly extends the effective obstacle area, enhancing the magnetic torque and promoting rapid migration.

To provide an upper bound of the intensity of magnetic interaction, we assume the dipolar magnetic topology. The properties of stellar wind are taken from the starAML polytropic magnetized wind models (\citealt{Reville}). The polytropic wind solutions reveal that, at close separations, the magnetic pressure dominates over the hydrodynamic pressure (\citealt{Reville2,Strugarek3}), which is why $p_\mathrm{tot} \approx \frac{B_\mathrm{*, loc}^2}{2 \mu_{0}}$ in our work, where $B_\mathrm{*, loc}$ is the stellar magnetic field at the planetary orbit. The magnetic field strength, density, and temperature at the base of the corona are obtained from eqs.(28)--(30) in \cite{Ahuir}.

The magnetic field strength of a planet, $B_\mathrm{pl}$, is one of the least known ingredients in the dipolar torque parametrization. The traditional dynamo scaling laws (\citealt{Russell,Stevenson,Sano}) imply that $B_\mathrm{pl}$ depends on the planetary spin (and hence on the orbital period in tidally-locked systems), resulting in relatively weak magnetic fields in the range of 0.1--10 G (\citealt{Zaghoo}). However, recent observational findings suggest it can reach the values of 10--100 G (\citealt{Cauley,Cauley1}), in agreement with the new scaling law based on the heat flux in the dynamo region (\citealt{Christensen,Davidson}). Here we consider that the flux is of internal origin and compute the grid of planetary models using MESA code and follow the approach described in \cite{Hori}. Fig.~\ref{fig1} demonstrates our estimates of the planetary equatorial field. 

\subsubsection{Mass loss contribution}

The planetary mass loss leads to the orbital expansion. However, the course of this process is determined by the fraction $\chi$ of the angular momentum of the outflow returned to the orbit:

\begin{equation}
     \left(\frac{\dot{a}}{a}\right)_\mathrm{ML} \approx -2 \chi \frac{\dot{M}_\mathrm{pl}}{M_\mathrm{pl}},
    \label{eq:mass}
\end{equation}

Throughout the whole paper, we imply $\dot{M}_\mathrm{pl}$ to be negative. The parameter $\chi$ incorporates the amount of the escaping mass remaining bound within a system after it flows out of a planet and the fraction of its angular momentum returned to an orbit via torques (it corresponds to $1-\gamma\sigma$ in a notation by \cite{Valsecchi1} and \cite{Jackson}, where $\gamma$ determines the amount of specific angular momentum comprised in the outflow and $\sigma$ is the fraction of an accretion disc escaping from a system). Similar to \cite{Jackson} and \cite{Fujita}, $\chi$ is fixed at 0.5 throughout this study except in subsection~\ref{subsec:chi} where the effects of different values are explored. We remind that adopting constant $\chi$ might be a crude assumption. In reality, we expect the fraction of returned angular momentum to be sensitive to the geometry of the evaporative wind, which is affected by various factors, such as stellar UV flux, escape velocity, orbital separation, and strength of the magnetic field (\citealt{Matsakos}). Finally, it is worth mentioning that $\chi$ must depend on whether a planet loses mass through photoevaporation or Roche-lobe overflow. For example, \cite{Valsecchi1} consider photoevaporation having a marginal impact on orbital migration. Though, the hydrodynamical simulations by \cite{Shaikhislamov} have shown that, for small orbital separations, the planetary wind total pressure is higher than the pressure of stellar wind all along the planet-star line, allowing ‘captured by the star’ regime to occur. Thereby, the escaping material might form a torus-like structure, exchanging its orbital momentum with a planet via tidal torques. Stellar irradiation is not powerful enough to disrupt the outflow, as reported by \cite{Debrecht}, which justifies including the feedback of the evaporated mass on the planetary orbit in our model.

\subsection{Photoevaporation}
\label{subsec:phot}
Two regimes of photoevaporation are implemented, namely energy-limited and recombination-limited regimes. At low incident flux, most of the heating goes to $PdV$ work since gas advection balances photoionization. As the planet approaches the host star, the atmospheric temperature increases, enhancing the Ly$\alpha$ cooling. Eventually, the radiative processes begin to dominate over adiabatic cooling, which marks the transition to a recombination-limited regime, where the outflow becomes less efficient. The principles underlying energy-limited and recombination-limited regimes have been studied in \cite{Murray-Clay} and \cite{OwenJackson}. Recently, \cite{Caldiroli} performed hydrodynamic simulations of the thermospheres of hot gas planets, which enabled to calibrate the evaporation efficiency to calculate the outflow rate in the form:

\begin{equation}
    |\dot{M}_\mathrm{pl, PE}| = \eta \frac{3 F_\mathrm{XUV}}{4GK \rho_\mathrm{pl}}.
    \label{eq:mloss1}
\end{equation}
$G$ is the gravitational constant, $\rho_\mathrm{pl}$ is the mean planetary density, $\eta$ is the photoevaporation efficiency computed as in \cite{Caldiroli}. Here, the XUV flux, $F_\mathrm{XUV}$, is derived using the X-ray and the X-ray–-EUV relations from \cite{Johnstone}.The impact of the host star gravitational pull is manifested by the potential energy reduction factor $K < 1$ (\citealt{Erkaev}):
\begin{equation}
    K = 1 - \frac{3}{2\phi} + \frac{1}{2\phi^3}, \; \phi = \frac{R_\mathrm{Rl}}{R_\mathrm{pl}}, 
    \label{eq:K}
\end{equation}
with the Roche lobe radius $R_\mathrm{Rl} = \frac{a}{f_\mathrm{p}} \left(\frac{M_\mathrm{pl}}{M_*}\right)^{1/3}$. The value of $f_\mathrm{p}$ depends on the planetary structure and constitution. In our reference model, we set $f_\mathrm{p} = 3^{1/3}$. In subsections~\ref{subsec:rlim} and \ref{subsec:model2}, we also run our simulations with $f_\mathrm{p} = 2.7$. 

 Note that $K$ tends to zero when the planet gets close to filling the Roche lobe, which makes eq.(\ref{eq:mloss1}) inappropriate for modeling the relevant cases. In order to solve this issue, we refer to the study by \cite{Murray-Clay} and the recombination-limited mass flux given by:
\begin{equation}
    |\dot{M}_\mathrm{pl, PE}| = 4\pi\rho_\mathrm{s}c_\mathrm{s}r_\mathrm{s}^2,
    \label{eq:mloss2}
\end{equation}
the subscript 's' denotes the sonic point. The calculation of $r_\mathrm{s}$ and $\rho_\mathrm{s}$ is adopted from \cite{Fujita}. The only significant change concerns the location of the photoionization base with respect to planetary radius $\beta =  R_\mathrm{XUV}/R_\mathrm{pl}$. In the former study, $\beta$ is assumed to be unity, while we derive it by applying the prescription from \cite{Salz}. To avoid unphysically large photoevaporation rates arising from the reduction term $K$ being close to zero, we select the minimum of the two values provided by eqs.(\ref{eq:mloss1}) and (\ref{eq:mloss2}). Finally, we recall that planetary magnetic field and stellar wind may substantially alter the dynamics of thermally driven outflow, as reported in \cite{Trammell} and \cite{Vidotto}. However, this effect is out of the scope of the present paper, and we leave it for future studies.
\subsection{Roche-lobe overflow}
\label{subsec:rlo}
The Roche-lobe overflow (RLO) occurs when a planet reaches the Roche limit, $a_\mathrm{R}$, defined as:
\begin{equation}
    a_\mathrm{R} = f_\mathrm{p} R_\mathrm{pl} \left({\frac{M_\mathrm{*}}{M_\mathrm{pl}}}\right)^{1/3}.
    \label{eq:rlimit}
\end{equation}

Throughout the RLO phase, the secular dynamics is governed by the condition:
\begin{equation}
    R_\mathrm{pl} = R_\mathrm{Rl}. 
    \label{eq:RLO}
\end{equation}
Differentiating eqs.(\ref{eq:RLO}) with the definition of the Roche lobe radius, leads to the following expression:
\begin{equation}
    \xi \frac{\dot{M}_\mathrm{pl, RLO}}{M_\mathrm{pl}} +  \kappa \frac{\dot{F}}{F}=  \frac{\dot{a}}{a} + \frac{1}{3}\frac{\dot{M}_\mathrm{pl, RLO}}{M_\mathrm{pl}},
    \label{eq:RLO1}
\end{equation}
where $\xi = \frac{\partial \ln{R_\mathrm{pl}}}{\partial \ln{\,M_\mathrm{pl}}}$ and $\kappa = \frac{\partial \ln{R_\mathrm{pl}}}{\partial \ln{F}}$. $F$ is the incident flux on a planet. After substituting the derivative of the orbital semi-major axis, given in subsection~\ref{subsec:migration}, and the derivative of the incident flux, we obtain:
\begin{equation}
    \frac{\dot{M}_\mathrm{pl, RLO}}{M_\mathrm{pl}} =  \frac{\left(\frac{\dot{a}}{a} \right)_\mathrm{tm}\left(2\kappa + 1\right) - \kappa\left(\frac{\dot{L}}{L}\right)}{\xi - \frac{1}{3} + 2\chi\left(2\kappa + 1\right)},
    \label{eq:RLO2}
\end{equation}
 $L$ is the luminosity of a star. Note that eq.(\ref{eq:RLO2}) provides the mass-loss rate required for the companion to stay on the Roche limit under the assumption that stable mass transfer is sustained. However, when the denominator in eq.(\ref{eq:RLO2}) becomes negative, it is no longer achievable, and the planetary evolution eventually ends up with a tidal disruption event. In particular, the unstable mass transfer may occur when the parameter $\chi$ is close to zero, implying the feedback of the escaping material on the planetary orbit is small. For a Jupiter-mass planet, the mass--radius dependence is weak, and in the zeroth approximation, one can assume $\xi = 0$. Furthermore, if we neglect the planetary inflation due to heating by setting $\kappa = 0$, we obtain the following condition for stable mass transfer  $\chi > 0.17$. Because of a small negative slope of the mass--radius relation for Jupiters with $M_\mathrm{pl} > 0.5 \; M_\mathrm{J}$ (see subsection~\ref{subsec:radius}), the instability region extends up to $\chi \sim 0.2$. Accordingly, adopting $\chi = 0.5$ yields stable mass transfer for all the cases involving RLO (nevertheless, following \cite{Metzger}, we still assume tidal disruption if the planetary mean density is above the stellar mean density at the onset of RLO).
 
 Eq.(\ref{eq:RLO2}) provides the total mass-loss rate during the RLO phase, including photoevaporation. Our $\dot{M}_\mathrm{pl, RLO}$ is equivalent to $\dot{M}_\mathrm{pl, RLO} +\dot{M}_\mathrm{pl, PE}$ in a notation by \cite{Valsecchi1}, who separated the contribution of thermally driven outflow and RLO transfer, assuming zero feedback of evaporated mass on the planetary orbit. We do not differentiate between mass-loss processes in terms of their subsequent impact on the orbital migration. Thus, the material lost through heating returns the same fraction of angular momentum as the material lost through RLO, which is reflected in using the single $\chi$ value throughout the simulation. The value of $\dot{M}_\mathrm{pl, PE}$, calculated as described in subsection~\ref{subsec:phot}, provides the relative contribution of photoevaporation to the total outflow during RLO.  When photoevaporative losses surpass RLO mass-loss rate (i.e., when $|\dot{M}_\mathrm{pl, PE}| > |\dot{M}_\mathrm{pl, RLO}|$), the outward torque becomes too strong to keep the planet on the Roche limit, and the overflow phase terminates with the planet moving outside $a_\mathrm{R}$.

We do not simulate the RLO phase when gravity waves dissipate by the time a planet reaches the Roche limit. First, it is unclear what happens after the planetary mass drops below the critical value required for the wave breaking to continue. As discussed in L21, the critical mass $M_\mathrm{crit}(M_{*},a,t)$ is a sharply decreasing function of stellar age. Given that the dissipation of g-modes makes the dominant contribution to the planetary migration rate (and hence to the numerator of the right side of eq.(\ref{eq:RLO2})), the planetary mass may follow the variation of $M_\mathrm{crit}(M_{*},a,t)$ to remain on the edge of gravity wave damping. If so, the mass loss is intense but steady. If gravity waves continue to break after the mass of the planet becomes less than $M_\mathrm{crit}(M_{*},a,t)$, the gaseous envelope is likely to disappear within a very short timescale. Each of these two outcomes is characterized by an extremely rapid planetary mass decrease, turning a gas giant into a naked core shortly after the onset of RLO. Second, our parametrization of planetary radius (see subsection~\ref{subsec:radius}) takes into account the inflation due to heating, which is assumed to proceed instantly. In reality, the inflation timescale is expected to be short (\citealt{Thorngren}), but so is the timescale of the migration driven by the gravity wave dissipation (\citealt{Barker}). It is likely that some of the most massive planets do not adjust their radii on time and undergo direct impact or tidal disruption instead of stable RLO. 

Stable mass transfer does not begin when the Roche limit is below the stellar surface. The corresponding case is called direct impact. In addition, following \cite{Metzger}, we assume that if the planetary density is higher than the stellar mean density, RLO occurs too close to the stellar surface to allow the formation of the accretion disc, which is why the angular momentum of the outflow does not return to the orbit, resulting in the unstable mass transfer. This scenario is called tidal disruption.

\begin{figure}
	\includegraphics[width=\columnwidth]{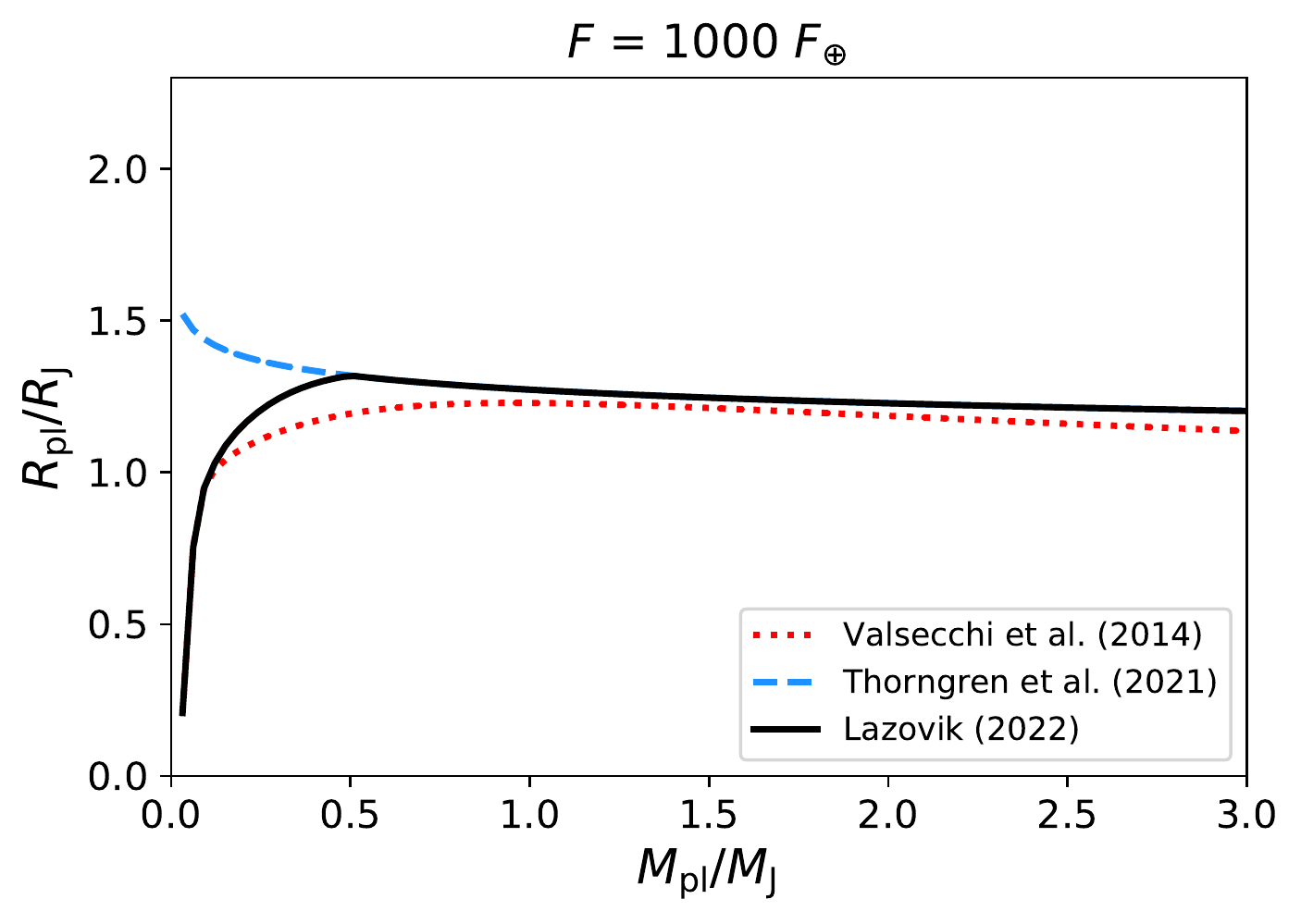}
	\includegraphics[width=\columnwidth]{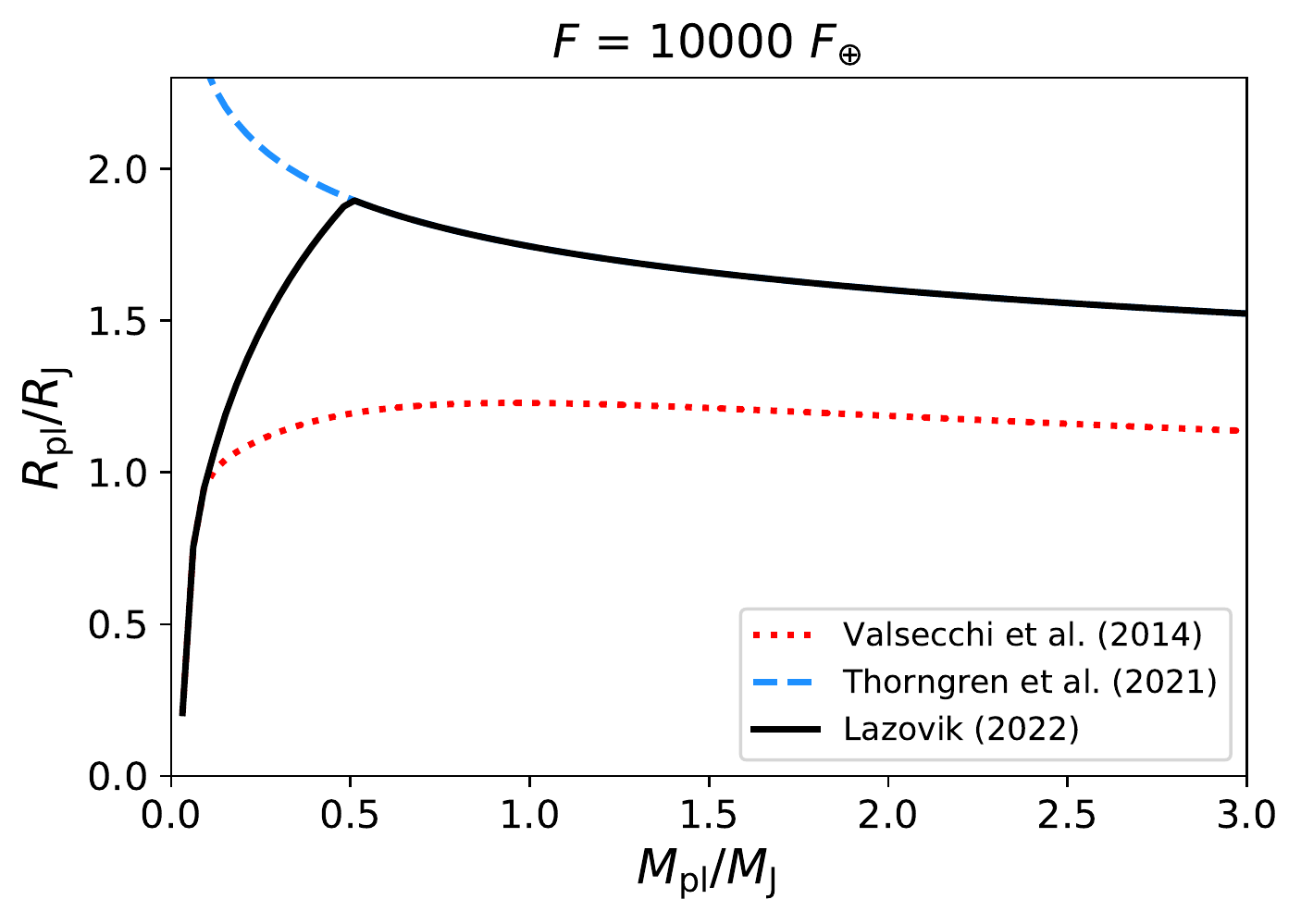}
    \caption{Planetary radius as a function of mass. Red dotted line represents the eq. (14) from~\protect\cite{Valsecchi}, blue dashed line corresponds to the scaling law from~\protect\cite{Thorngren1}, black solid line is the parametrization adopted in the present study.  Top panel refers to $F = 1000 \; F_{\oplus}$, bottom panel refers to $F = 10000 \; F_{\oplus}$.}
    \label{fig2}
\end{figure}

\subsection{Planetary radius}
\label{subsec:radius} 

To parametrize the planetary radius, we divide the planets into three groups depending on their mass. For planets with $M_\mathrm{pl} > 0.5\;M_\mathrm{J}$, we follow \cite{Thorngren}:
\begin{equation}
    R_\mathrm{pl} = R_\mathrm{pl,T} = A R_\mathrm{J} \;  \left({\frac{M_\mathrm{pl}}{M_\mathrm{J}}}\right)^{B}\;\left(\frac{F}{10^9}\right)^{C+D \; \log \;\frac{M_\mathrm{pl}}{M_\mathrm{J}}},\; \; M_\mathrm{pl} > 0.5\; M_\mathrm{J},
    \label{eq:radii1}
\end{equation}
where $F$ is in units of $\mathrm{erg \; s^{-1} \; cm^{-2}}$, $A = 1.22$, $B = -0.42$, $C = 0.137$, and $D = -0.072$. The corresponding empirical law is based on the calculated radii of the observational hot Jupiter sample and takes into account the planetary inflation under the effect of insolation.

The radii of planets with $M_\mathrm{pl} \leq 0.1\;M_\mathrm{J}$ are calculated according to \cite{Valsecchi}.  For this purpose, we adopt eq. (14) from the above paper, corresponding to the models with $M_\mathrm{c} = 10\;M_{\oplus}$:
\begin{equation}
\begin{aligned}
    R_\mathrm{pl} & = R_\mathrm{pl,V} = 0.67 R_\mathrm{J} \; \exp{\left(-4.9 \times 10^{-6}\times  \left({\frac{M_\mathrm{pl}}{M_\mathrm{J}}}\right)^{-4}\right)}\; + \\ 
    &+0.61 \; \left({\frac{M_\mathrm{pl}}{M_\mathrm{J}}}\right)^{1/3}\;\left(1 -  \exp{\left(-1.7 \times \left({\frac{M_\mathrm{pl}}{M_\mathrm{J}}}\right)^{-0.9}\right)}\right),  \; \;M_\mathrm{pl}  \leq 0.1\; M_\mathrm{J}.
    \end{aligned}
    \label{eq:radii2}
\end{equation}

For planets lying inside the [0.1, 0.5] $M_\mathrm{J}$ range, we calculate the radius by interpolating between the values obtained using eqs.(\ref{eq:radii1}) and (\ref{eq:radii2}):
\begin{equation}
    R_\mathrm{pl} = R_\mathrm{pl,T} + \left(R_\mathrm{pl,V} - R_\mathrm{pl,T} \right)\frac{0.5 - \frac{M_\mathrm{pl}}{M_\mathrm{J}}}{0.4}, \; \; 0.1\; M_\mathrm{J} < M_\mathrm{pl} \leq 0.5\; M_\mathrm{J},
    \label{eq:radii3}
\end{equation}

Such a form of dependence allows our estimates of the hot Jupiter radius to be consistent with the recent observational data. Besides, it enables us to study lower mass range characterized by a smaller inflation efficiency (\citealt{Thorngren1}). Fig.~\ref{fig2} shows the planetary radius plotted against mass for different incident fluxes.

Adopting the above relation implies that the planet is always in the state of thermal equilibrium, i.e., the adjustment of planetary radius to the mass loss and incident flux variation proceeds instantly. Generally, this is likely to be a reasonable assumption since \cite{Thorngren} reported that the time lag behind the equilibrium radius is too small to be inferred from the observations. However, in extreme cases, when the timescales of the orbital or mass evolution are small thermal equilibrium may be violated. Although we do not check this possibility, the conditions regarding thermal equilibrium deserve special attention in further research.

\section{Outline of secular evolution of hot Jupiter systems}
\label{sec:outline}

To infer the evolution of a star-planet system, we solve our differential equations governing the dynamics of the stellar rotation rate (eq. (\ref{eq:rot})), planetary semi-major axis (eq. (\ref{eq:migr})), and mass-loss rate (eqs. (\ref{eq:mloss1}), (\ref{eq:mloss2}), and (\ref{eq:RLO2})) using the Python SciPy routine odeint. The adopted time grid is composed of $2 \times 10^6$  points uniformly spaced between $\tau_\mathrm{disc}$ and ZAMS and $2 \times 10^6$ points uniformly spaced between ZAMS and zero age core helium burning. The relevant timesteps ($\sim 10$ yrs and  $\sim 10^4$ yrs in our calculations before and after ZAMS, respectively) are low enough to avoid convergence issues when the timescales associated with migration or mass-loss become low and to accurately capture the moment when a planet attaches the Roche limit or the $P_\mathrm{orb} = \frac{1}{2} P_\mathrm{rot}$ limit.

\begin{figure}
	\includegraphics[width=\columnwidth]{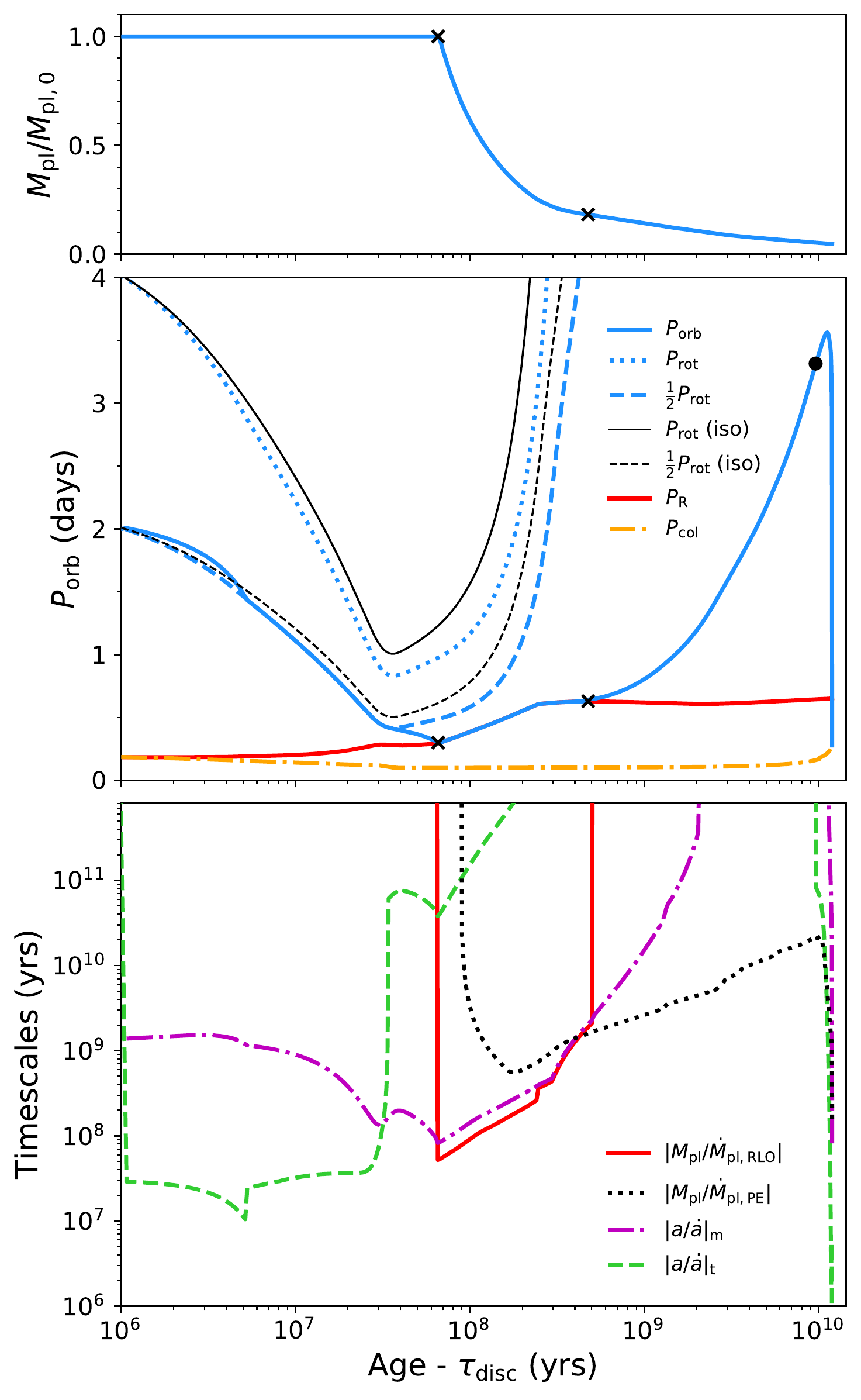}
    \caption{Secular evolution of hot Jupiter with $M_\mathrm{pl, 0} = 2\;M_\mathrm{J}$ around a solar-mass star with $P_\mathrm{rot,0}$ = 4.5 days. Top panel: planetary mass. Middle panel: orbital period. Blue solid line corresponds to the orbital period of a planet. Black solid and dashed lines indicate the synchronization period ($P_\mathrm{orb} = P_\mathrm{rot}$) and the inertial wave excitation ($P_\mathrm{orb} = \frac{1}{2} P_\mathrm{rot}$) limit of an isolated star, respectively. Blue dotted and dashed lines represent the synchronization period and the inertial wave excitation limit of a star in the presence of a planet, respectively. Yellow dash-dotted line represents the orbital period at which the collision with the host occurs $P_\mathrm{col}$. Red solid line indicates the Roche period $P_\mathrm{R}$. Black crosses mark the start and the end of the RLO phase. Black circle indicates the onset of gravity wave dissipation. Bottom panel: evolution of the relevant timescales. Green dashed and magenta dash-dotted lines represent migration due to tidal and magnetic interaction, respectively. Red solid and black dotted lines denote mass loss due to RLO and photoevaporation, respectively. Note that the migration timescale associated with mass loss is represented by the lowest of the latter two curves at any given time.}
    \label{fig3}
\end{figure}
\subsection{Reference RLO cases}
\label{subsec:ref}

\begin{figure*}
\begin{multicols}{2}
    \includegraphics[width=\linewidth]{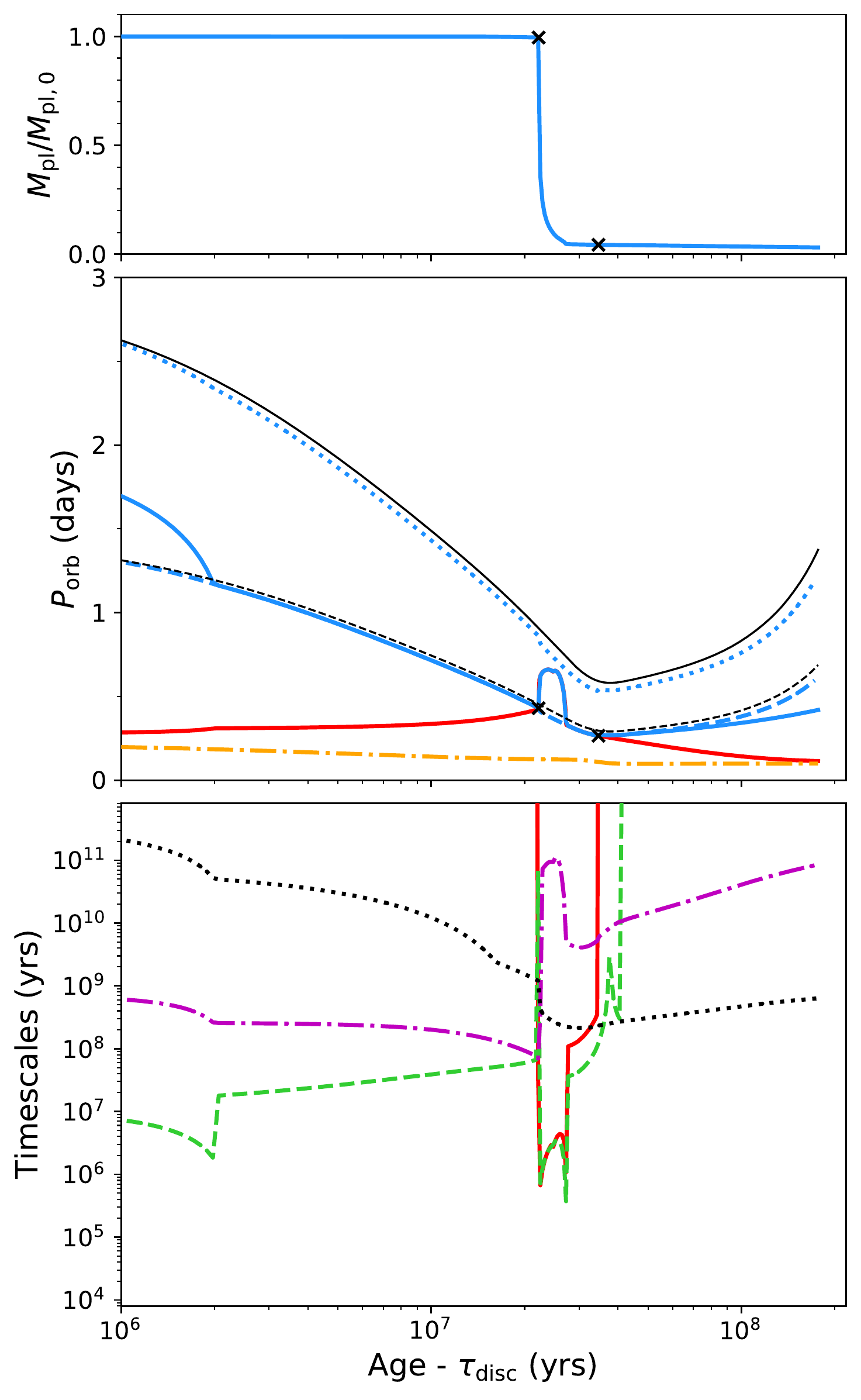}\par 
    \includegraphics[width=\linewidth]{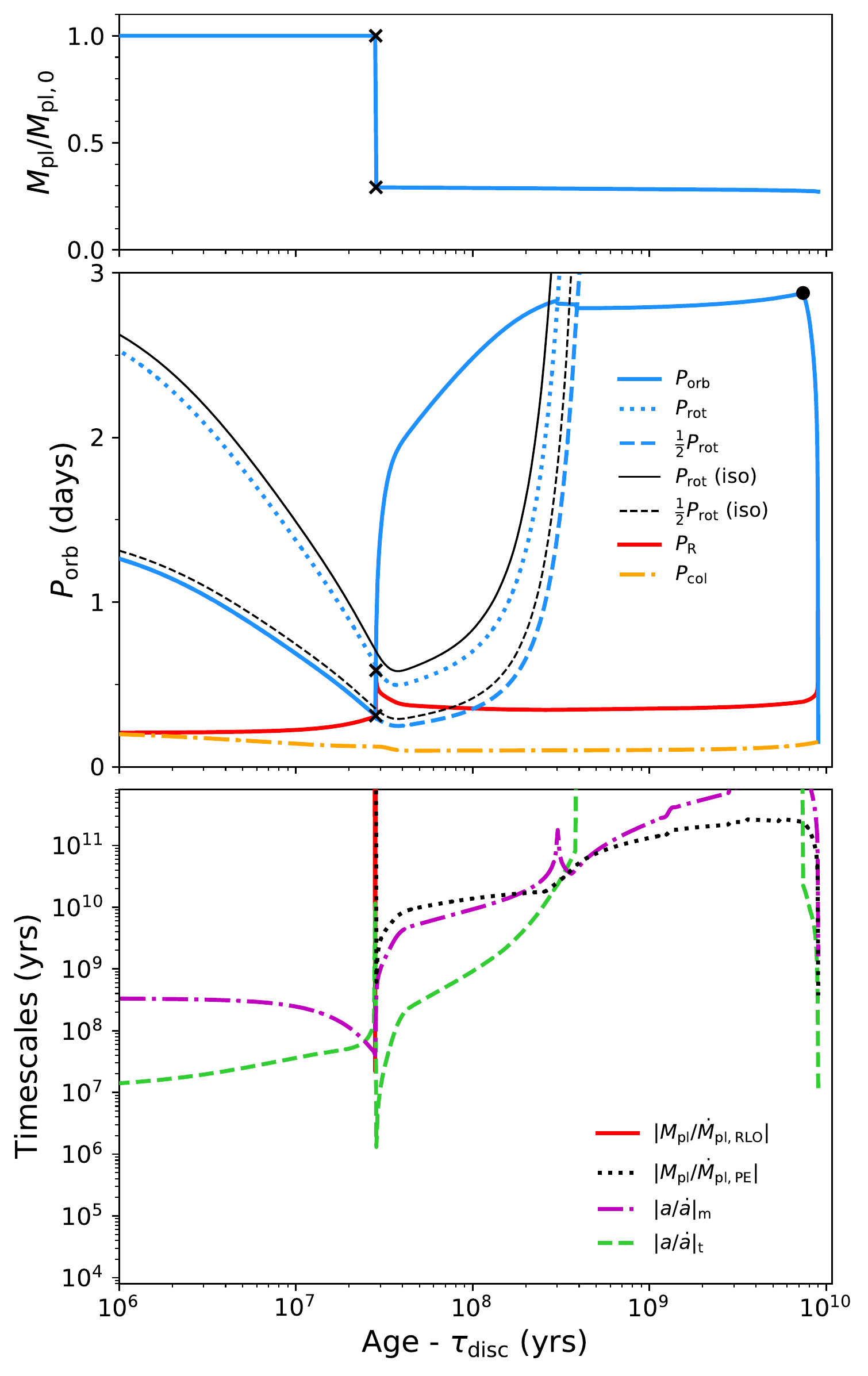}\par  
    \end{multicols}
\caption{Same as Fig.~\ref{fig3}, but for $M_\mathrm{pl, 0} = 1\;M_\mathrm{J}$, $P_\mathrm{rot,0}$ = 3 days (left panel) and $M_\mathrm{pl, 0} = 2\;M_\mathrm{J}$, $P_\mathrm{rot,0}$ = 3 days (right panel).}
\label{fig4}
\end{figure*}

Before moving to the main features characterizing the dynamics of hot Jupiters within our model, we focus on specific examples of secular evolution encompassing RLO. Fig.~\ref{fig3} demonstrates the reference case when hot Jupiter encounters the Roche limit in the MS stage. The corresponding star-planet system is composed of a solar-mass star with a median initial rotation ($P_\mathrm{rot,0}$ = 4.5 days) and hot Jupiter with $M_\mathrm{pl,0} = 2\;M_\mathrm{J}$ and $P_\mathrm{orb,0}$ = 2 days. During the pre-main sequence (PMS) stage, the dissipation of inertial waves drives rapid migration as the planet gets captured by the $P_\mathrm{orb} = \frac{1}{2} P_\mathrm{rot}$ limit, under which inertial waves are no longer excited. This limit is shown in the middle panel by the blue dashed line. As long as the host star spins up, the reference planet cannot leave the $P_\mathrm{orb} = \frac{1}{2} P_\mathrm{rot}$ limit. Below, the equilibrium tide dissipation is too weak to provide the migration rates required to approach the stellar surface without being captured by the limit again. On the contrary, above the edge of inertial wave excitation, tidal dissipation is too strong, which is why the planet does not move beyond the limit either. During the PMS phase of the orbital evolution, hot Jupiter rapidly oscillates around the $P_\mathrm{orb} = \frac{1}{2} P_\mathrm{rot}$ limit, and so does the tidal migration timescale. Though, we do not show these oscillations in the bottom panel of fig.~\ref{fig3}, where the relevant timescales are displayed. Instead, we illustrate the averaged tidal migration timescale around which the oscillations occur. The corresponding tidal migration rate is equivalent to the one required to stay on the $P_\mathrm{orb} = \frac{1}{2} P_\mathrm{rot}$ limit. By the beginning of the MS, when stellar contraction stops, hot Jupiter enters the region where the magnetic interaction provides the main contribution to the migration, allowing the planet to reach the Roche limit (denoted by the Roche period $P_\mathrm{R} = 2\pi \sqrt{\frac{a_\mathrm{R}^3}{GM_*}}$, shown in red) within several tens of Myr. Before the RLO phase, restricted by black crosses, the planet does not lose mass through photoevaporation since its gravitational potential is too high to allow atmospheric escape (\citealt{Caldiroli}). At the Roche limit, hot Jupiter moves away from the host within the first 200 Myr, as its current mass is in the plateau of Fig.~\ref{fig2}. The outward migration quenches when the planetary radius begins to shrink at $M_\mathrm{pl} \sim 0.5 \; M_\mathrm{J}$. The stable mass transfer lasts for almost 1 Gyr, over which the planet loses most of its gaseous envelope and reaches the bottom limit of the Jovian planet mass range. By that time, the mass-loss rate estimated from eq.(\ref{eq:RLO2}) is significantly reduced as the planetary mass and radius decrease, and so is the planetary magnetic field strength. Even though the photoevaporation rate also tends to slow down since the star produces less high-energy radiation, the outflow driven by the atmospheric escape starts to prevail over the RLO losses. This is shown in the bottom panel, which compares the timescales associated with the orbital and mass evolution of the corresponding system. The planet undergoes outward migration and eventually becomes hot Neptune, which remains stable against tidal inspiral until the MS termination. The rapid expansion of the evolved star and the accompanying enhancement of equilibrium tide dissipation efficiency result in planetary ingestion at $P_\mathrm{orb} = P_\mathrm{col} \equiv 2\pi \sqrt{\frac{(R_{*} + R_\mathrm{pl})^3}{GM_*}}$ (the orbital period at which the collision occurs, $P_\mathrm{col}$, is represented by yellow dash-dotted line). 

In turn, it is unclear what happens if the planet fills the Roche lobe before the host star has reached the MS. As noted in L21, the planets around fast rotators are subject to intense tidal interaction driven by inertial waves at the early ages of stellar evolution. Above the $P_\mathrm{orb} = \frac{1}{2} P_\mathrm{rot}$ limit, the migration timescale can be shorter than 1 Myr. Thus, the stability of mass transfer under such extreme conditions is questionable. If mass transfer remains stable, the outcome of the orbital evolution primarily depends on whether the Roche limit can reach the corotation radius of the contracting star. If the Roche limit does not encounter the corotation radius, the planet moves back to the $P_\mathrm{orb} = \frac{1}{2} P_\mathrm{rot}$ limit and continues losing mass via RLO until photoevaporation starts to dominate. This scenario is demonstrated on the example of hot Jupiter with $M_\mathrm{pl, 0} = 1\;M_\mathrm{J}$ in the left panel of Fig.~\ref{fig4}. In the end, the corresponding planet loses all its gaseous envelope within 200 Myr. In the opposite scenario, shown in the right panel, hot Jupiter with $M_\mathrm{pl, 0} = 2\;M_\mathrm{J}$ crosses the corotation radius and enters the region where tidal and magnetic interactions promote orbital expansion, and RLO is no longer feasible. Thereby, the RLO phase lasts $7.5 \times 10^4$ yrs, much shorter than the case demonstrated on the left. Because the tidal migration timescale drops below $10^5$ yrs, over half the planetary mass is lost within this short interval. Throughout the MS, the host star spins down due to magnetic braking, and the outward migration halts at $P_\mathrm{orb}$ = 2.8 days, as the planet returns inside the corotation radius. At the end of the MS lifetime, the gravity waves begin to overturn the background stratification and break in the stellar interior, promoting rapid migration. Hot Jupiter eventually spirals down within 1 Gyr.

\begin{figure*}
\begin{multicols}{2}
	\includegraphics[width=\linewidth]{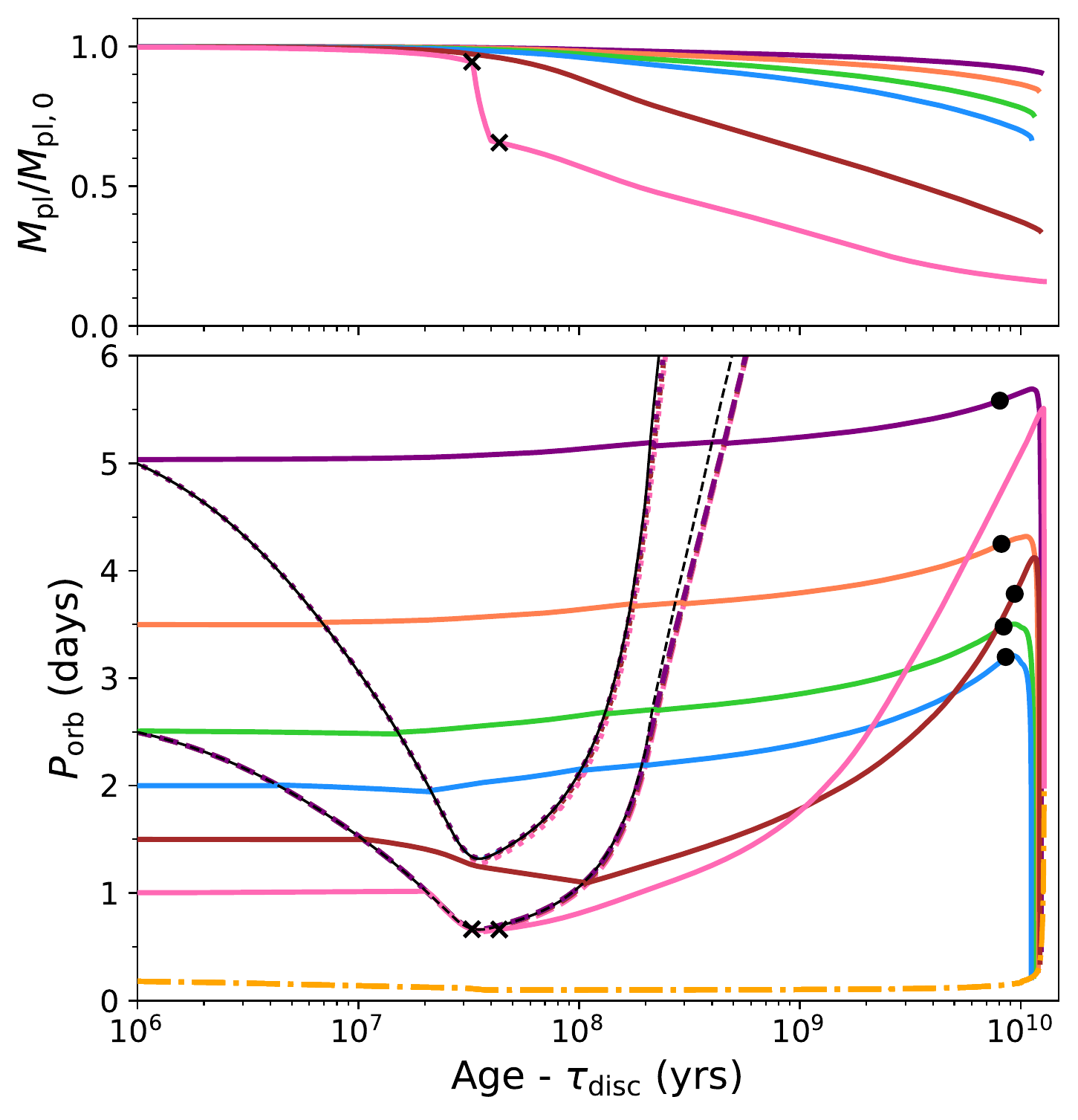}\par 
	\includegraphics[width=\columnwidth]{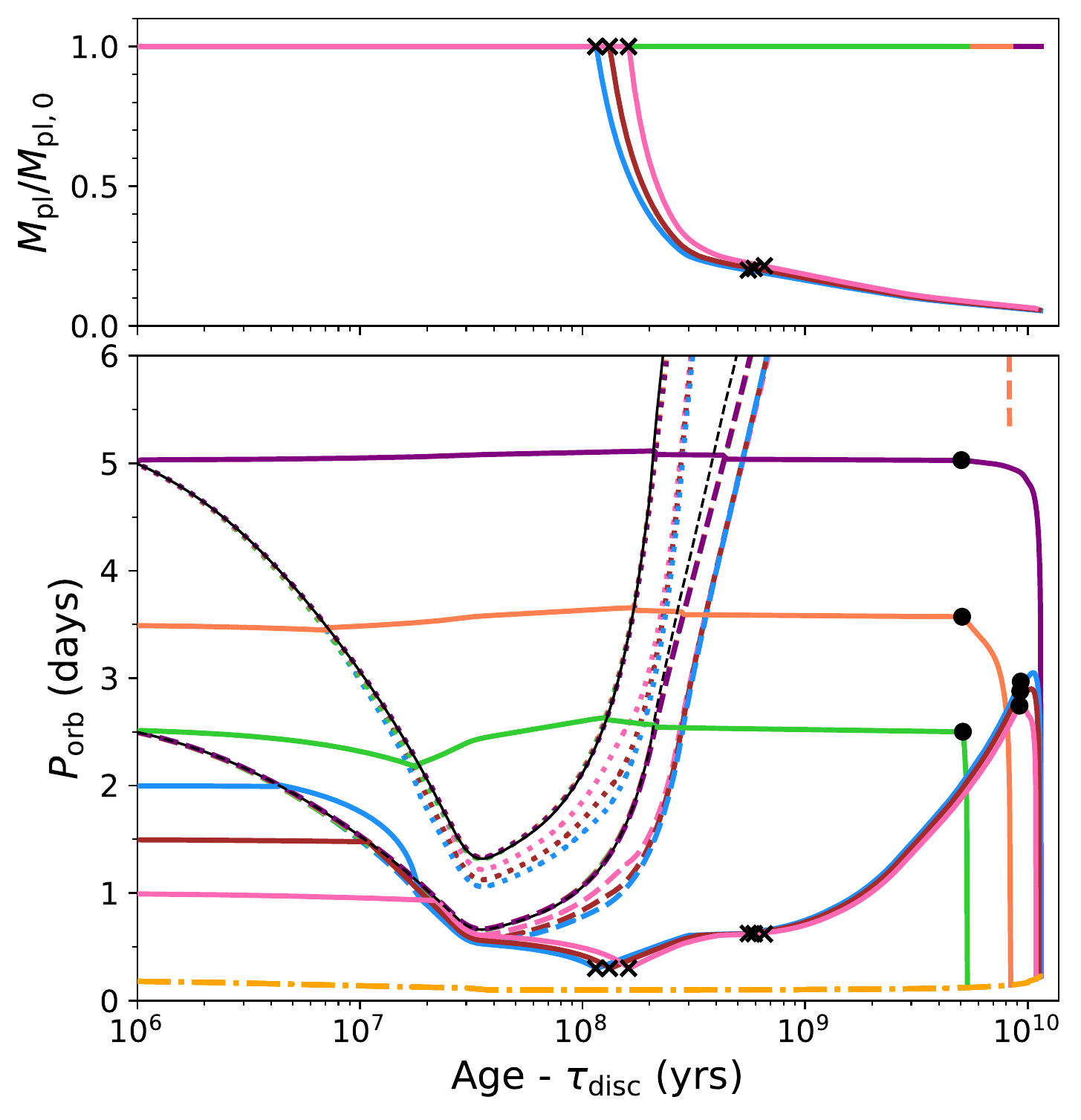}\par
    \end{multicols}
        \caption{Same as Fig.~\ref{fig3}, but for $M_\mathrm{pl, 0} = 0.3 \;M_\mathrm{J}$, $P_\mathrm{rot,0}$ = 5.5 days (left panel) and $M_\mathrm{pl, 0} = 2\;M_\mathrm{J}$, $P_\mathrm{rot,0}$ = 5.5 days (right panel). From bottom to top: $P_\mathrm{orb,0}$ = 1, 1.5, 2, 2.5, 3.5, and 5 days.}
\label{fig5}
\end{figure*}

\subsection{Impact of the initial semi-major axis and planetary mass}
\label{subsec:axmass}
We now explore the influence of the key model parameters on the secular evolution of a star-planet system. Among them, the initial planetary location and mass play a substantial role, as shown in Fig.~\ref{fig4}. Note that hot Jupiters located far enough from the moderately rotating host do not undergo significant migration until the onset of gravity wave damping, represented by black circles. For more massive planets, gravity wave breaking begins earlier, and the associated dissipation rate is higher, which increases the probability of infall before the terminal-age main sequence (TAMS). Lower mass planets, shown in the left panel, are affected by photoevaporation, driving mild orbital expansion.  Only the closest hot Jupiter with $P_\mathrm{rot,0}$ = 1 day reaches the Roche limit, initiating the stable mass transfer lasting for 10 Myr. In contrast, higher-mass planets are more likely to proceed through RLO. However, given that the Roche limit is closer to the host, it takes more time for these hot Jupiters to encounter it. One can also notice that, among the planets undergoing RLO, initially the further ones fill the Roche lobe earlier. This feature is due to the angular momentum exchange, which leads to stellar spin-up and the $P_\mathrm{orb} = \frac{1}{2} P_\mathrm{rot}$ limit lowering, allowing the system with the higher angular momentum of the planetary orbit to shrink to a smaller separation by ZAMS.

Post-RLO phase of the orbital evolution also depends on the initial planetary mass. As noted in subsection~\ref{subsec:rlo}, RLO stops when the photoevaporative mass-loss rate exceeds the mass loss required for the planet to stay on the Roche limit. The mass-loss rate during RLO is determined by the planetary magnetic field since, in the absence of the dynamical wave dissipation, the magnetic forces prevail over the tidal forces, adjusting the mass transfer. We demonstrated in Fig.~\ref{fig1} that, after the emergence of the dynamo region, the magnetic field strength is decreasing function of age and mass. Consequently, the late onset of RLO results in the planet being able to sustain a higher fraction of its gaseous envelope by the time when thermally driven outflow starts to dominate. On the contrary, less massive components are more prone to active photoevaporation, and the early finish of RLO gives more time for the orbital expansion, which explains why the initially shortest-period lower-mass planet in the left panel of Fig.~\ref{fig5} migrates further away from the host than the higher-mass planet undergoing RLO in the right panel.

\begin{figure*}
\begin{multicols}{2}
	\includegraphics[width=\linewidth]{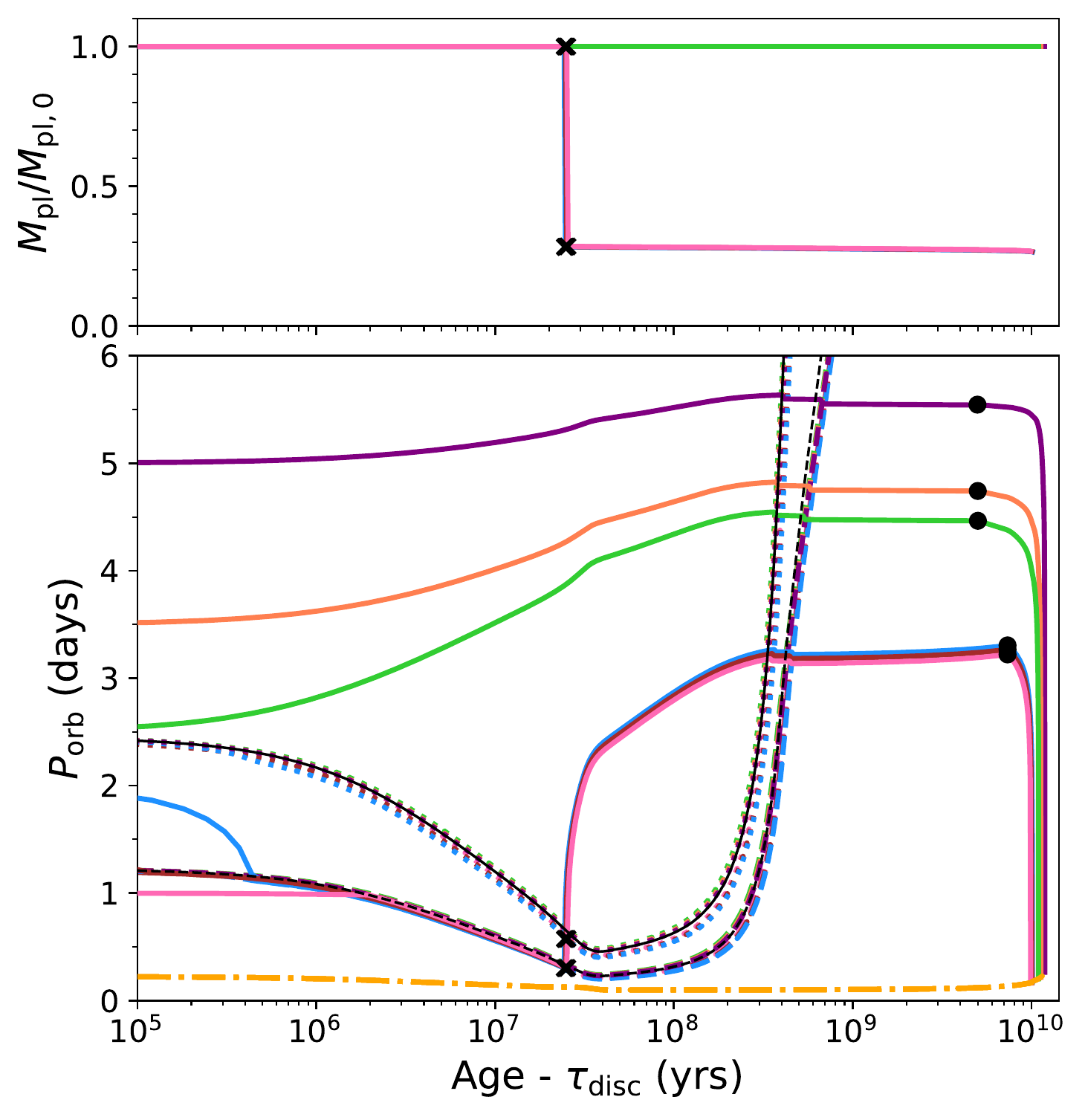}\par 
	\includegraphics[width=\columnwidth]{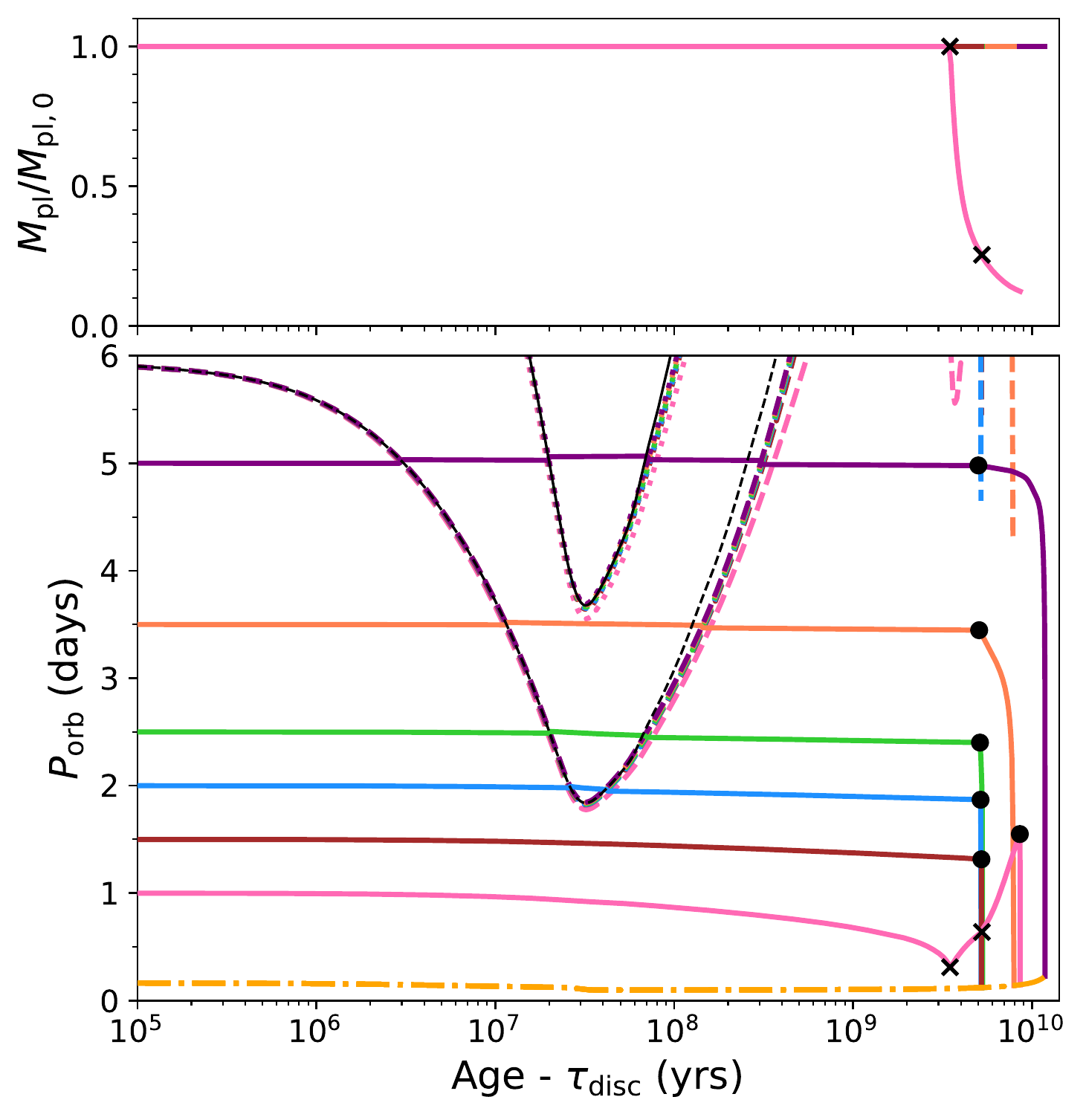}\par
    \end{multicols}
        \caption{Same as Fig.~\ref{fig5}, but for $M_\mathrm{pl, 0} = 2\;M_\mathrm{J}$, $P_\mathrm{rot,0}$ = 2.5 days (left panel) and $M_\mathrm{pl, 0} = 2\;M_\mathrm{J}$, $P_\mathrm{rot,0}$ = 12 days (right panel).}
\label{fig6}
\end{figure*}

\subsection{Impact of the initial stellar rotation rate}
\label{subsec:spin}

\begin{figure*}
\begin{multicols}{2}
    \includegraphics[width=\linewidth,height=5.4cm]{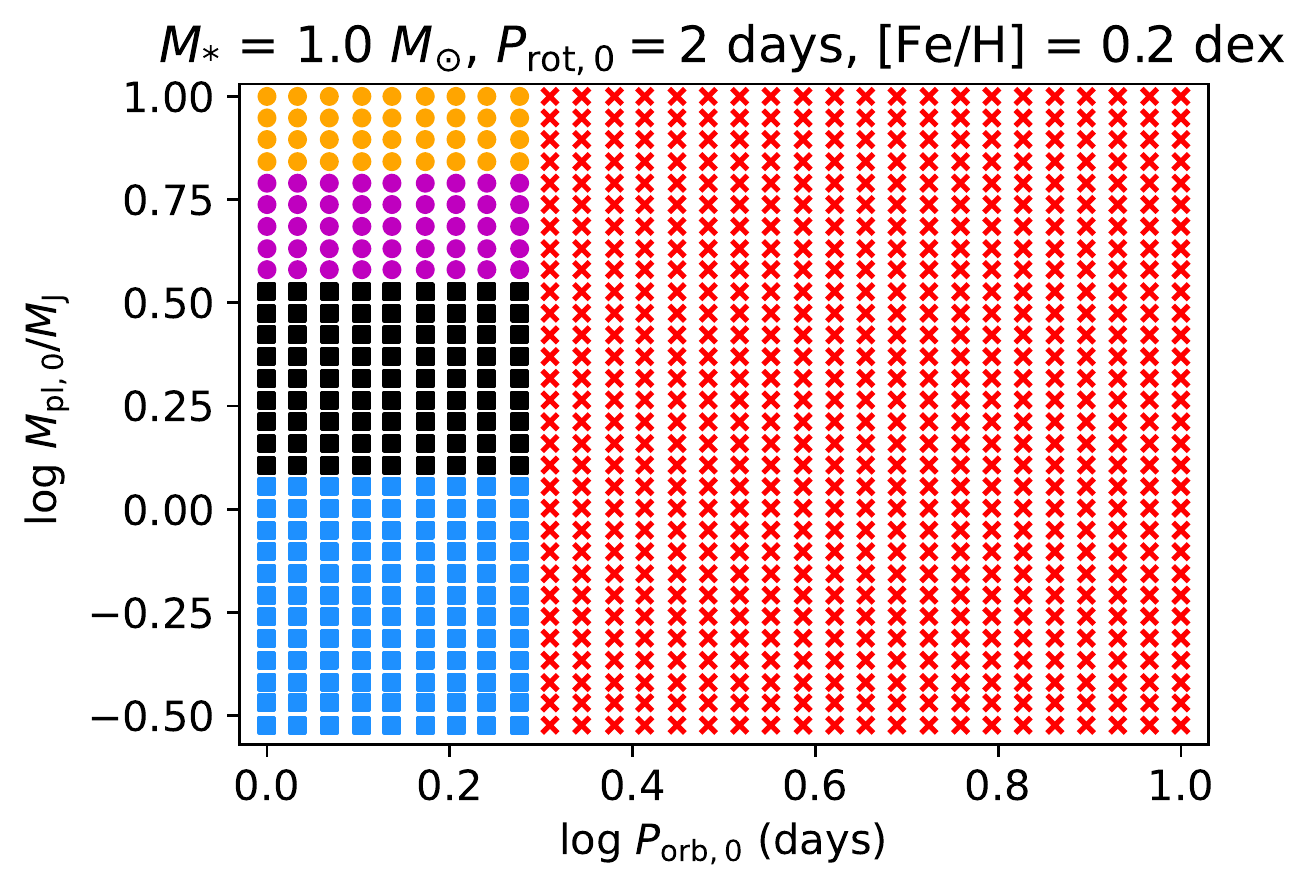}\par 
    \includegraphics[width=\linewidth,height=5.4cm]{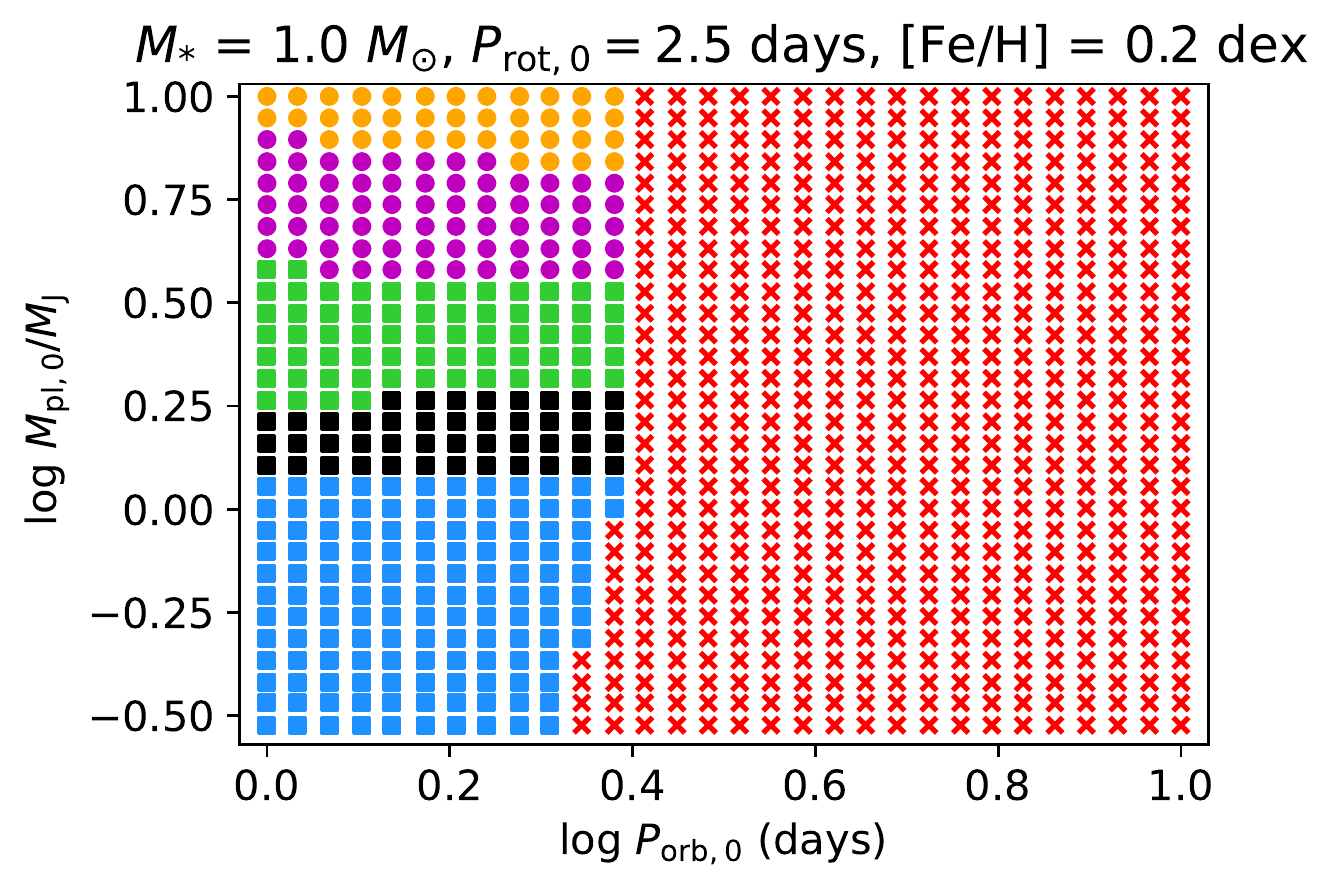}\par
    \includegraphics[width=\linewidth,height=5.4cm]{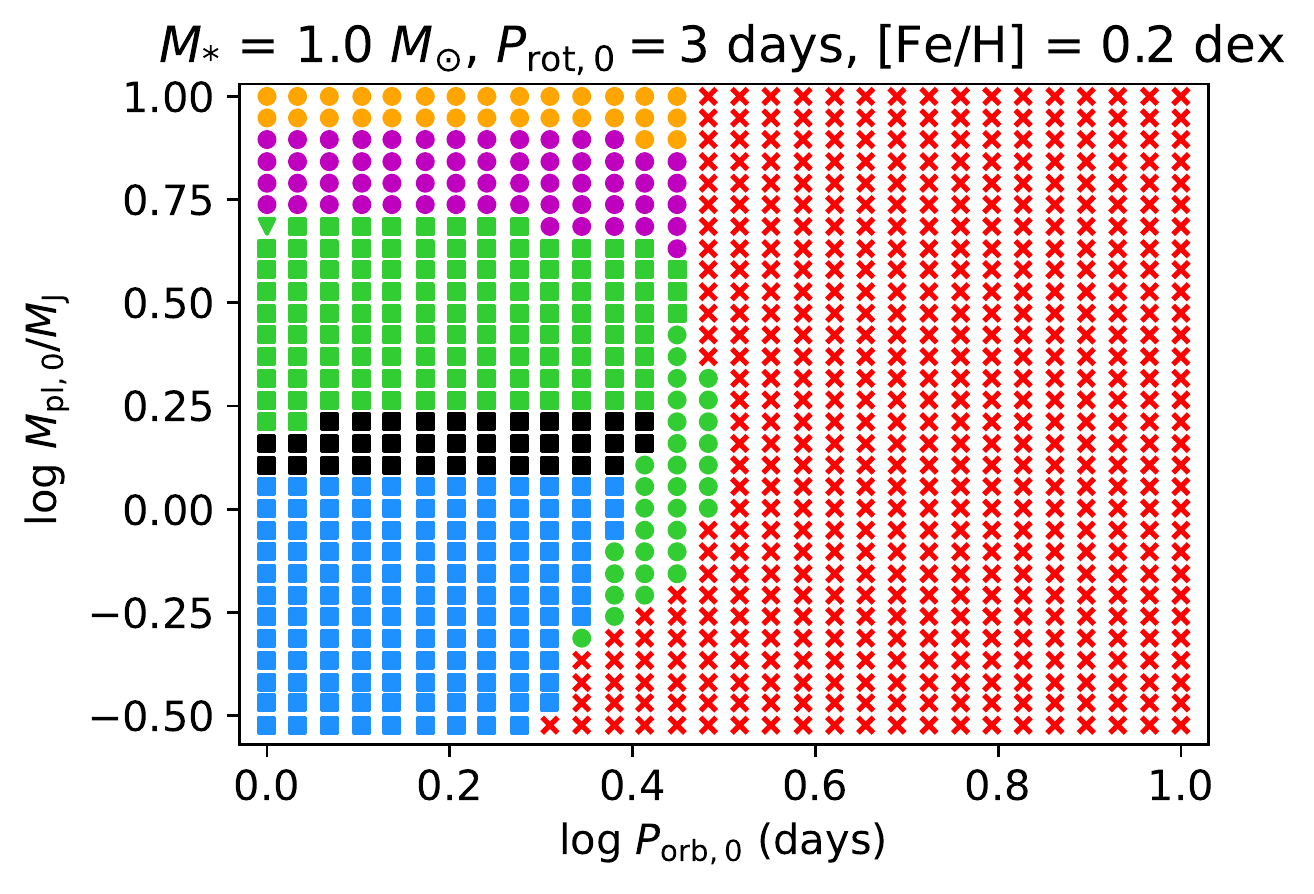}\par 
    \includegraphics[width=\linewidth,height=5.4cm]{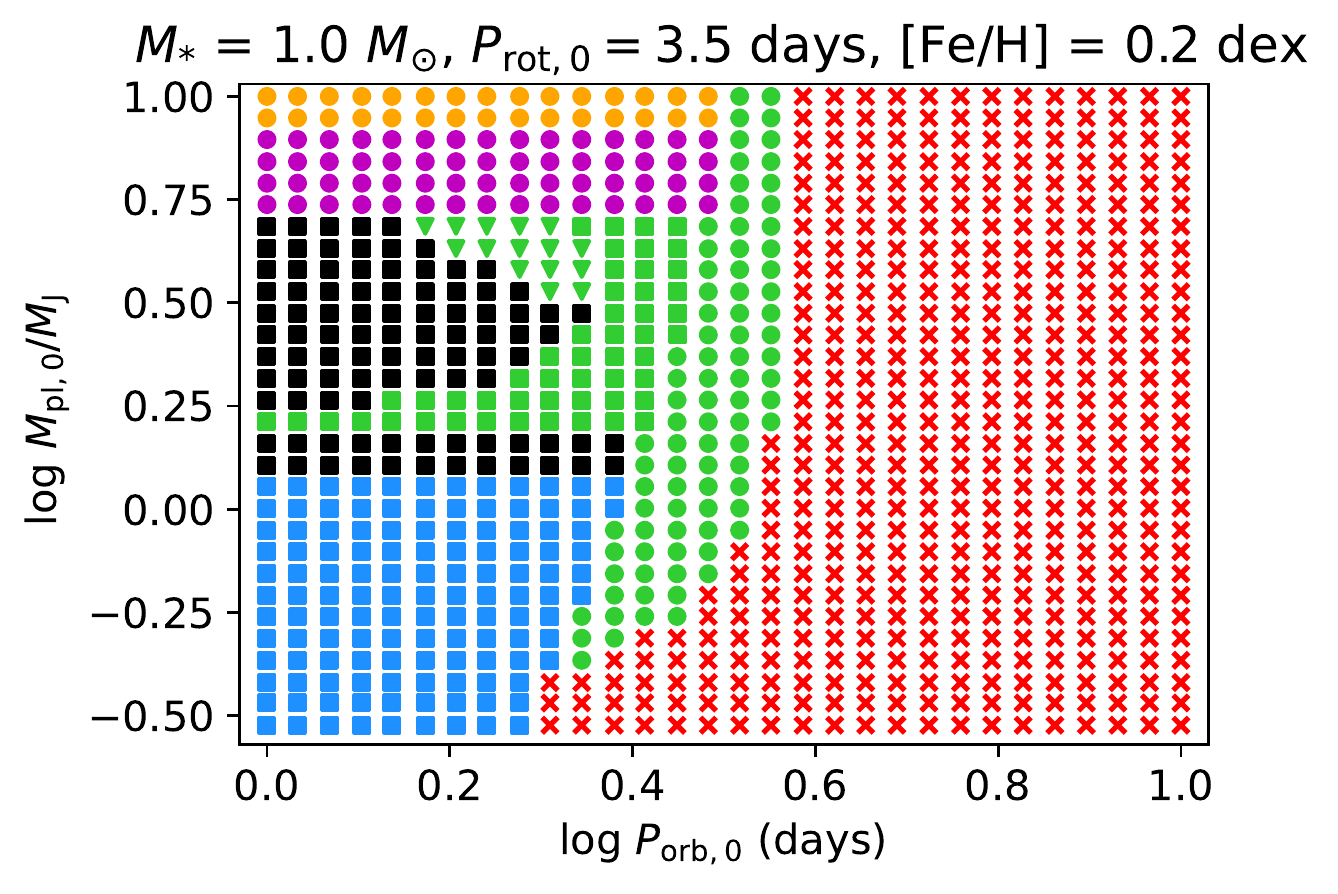}\par
    \includegraphics[width=\linewidth,height=5.4cm]{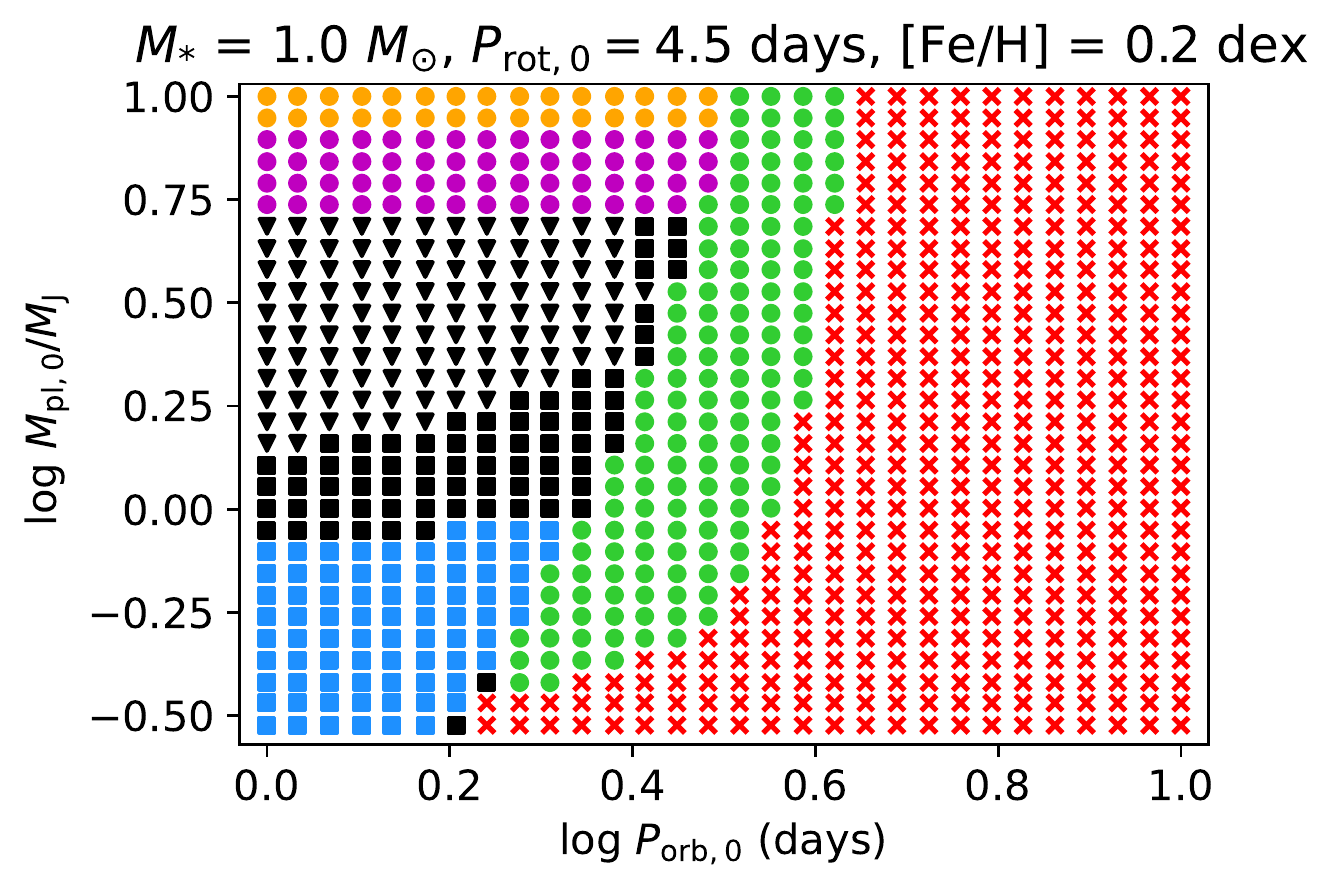}\par 
    \includegraphics[width=\linewidth,height=5.4cm]{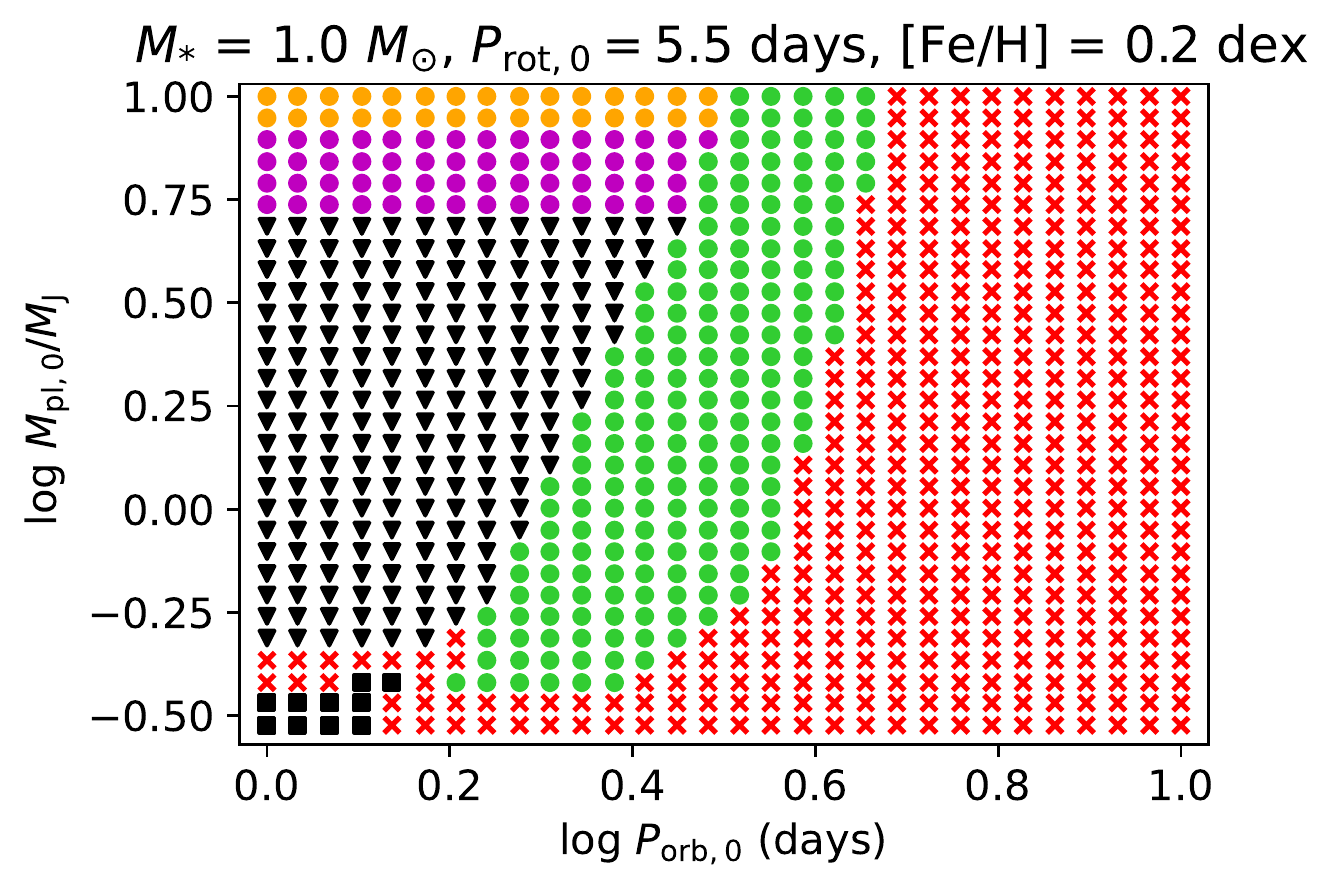}\par
    \includegraphics[width=\linewidth,height=5.4cm]{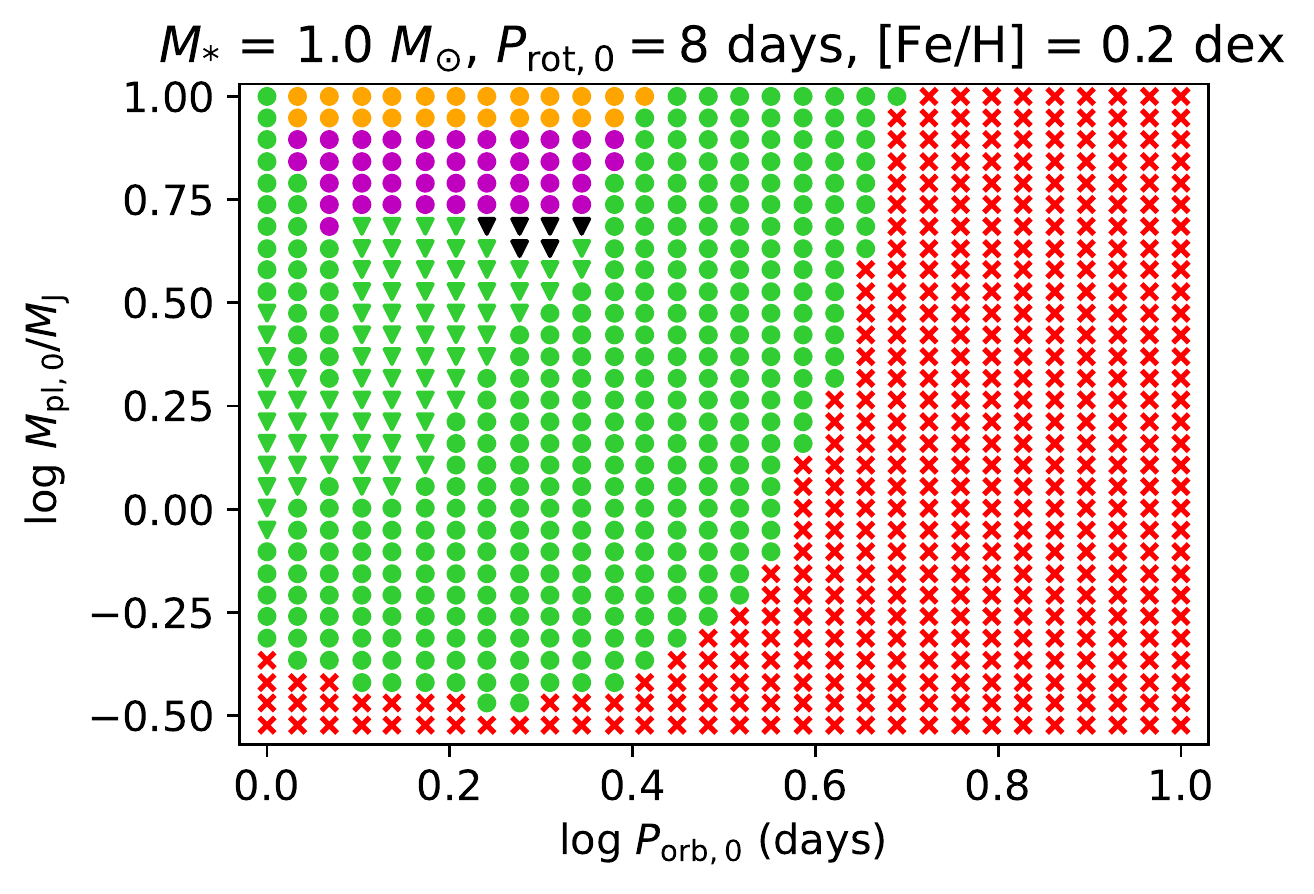}\par 
    \includegraphics[width=\linewidth,height=5.4cm]{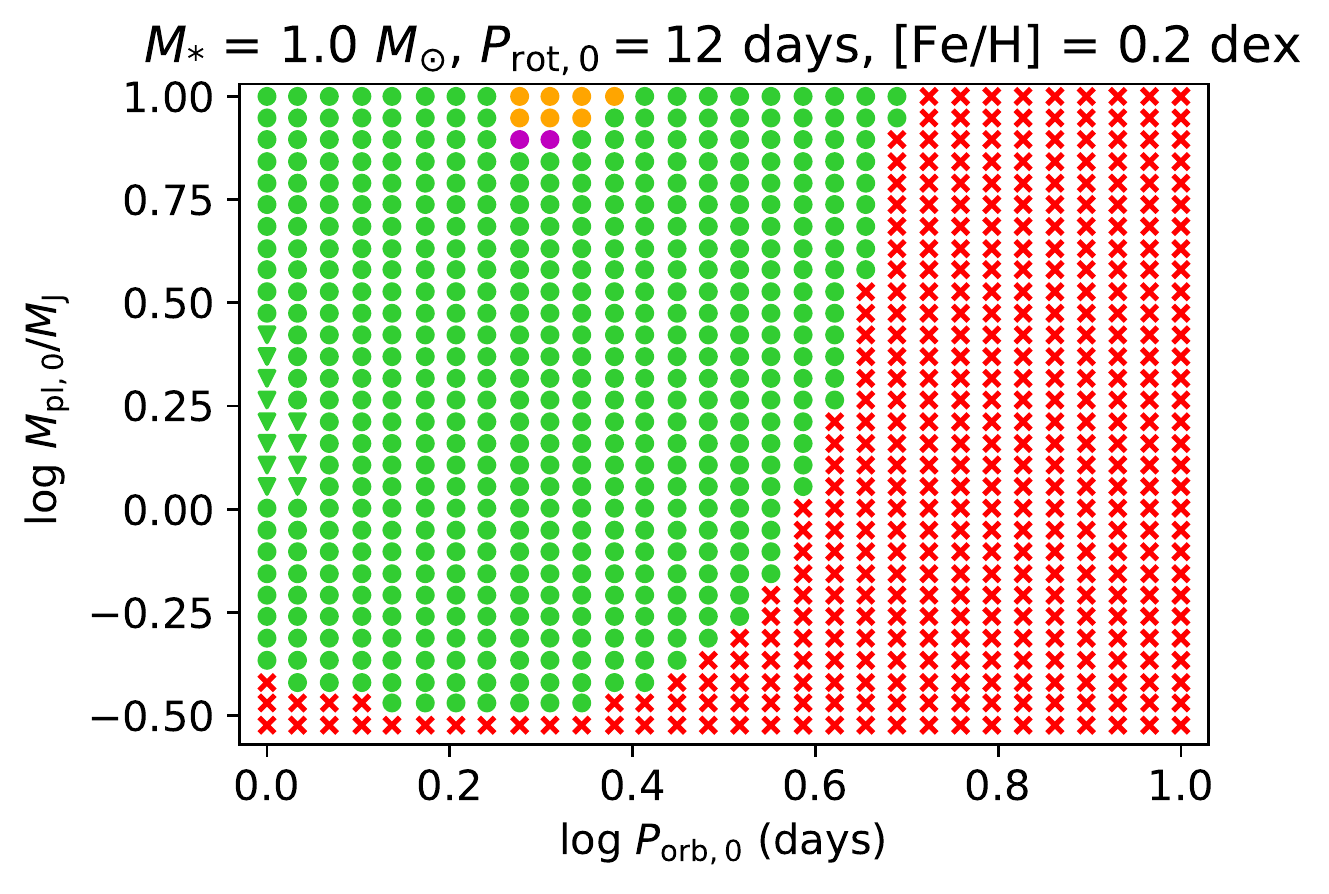}\par
    \end{multicols}
    \vspace*{-6mm}
\caption{Event diagrams for a solar-mass star with [Fe/H] = +0.2 dex and different initial rotation rates. The depicted merger events are prior to the end of the main sequence of the host. Red crosses: planets that do not undergo stable mass transfer, tidal disruption, or direct impact. Green color: mergers due to gravity wave dissipation. Yellow and purple colors: mergers before the initiation of gravity wave dissipation (direct impact and tidal disruption events, respectively).  Blue color: complete evaporation. Black color: planets that remain stable and keep the fraction of its gaseous envelope above 1 \% of the total planetary mass by TAMS. Squares and triangles: the cases when RLO starts before and after ZAMS, respectively. Circles: no RLO.}
\label{ap1}

\end{figure*}

\begin{figure*}
\begin{multicols}{2}
	\includegraphics[width=\linewidth]{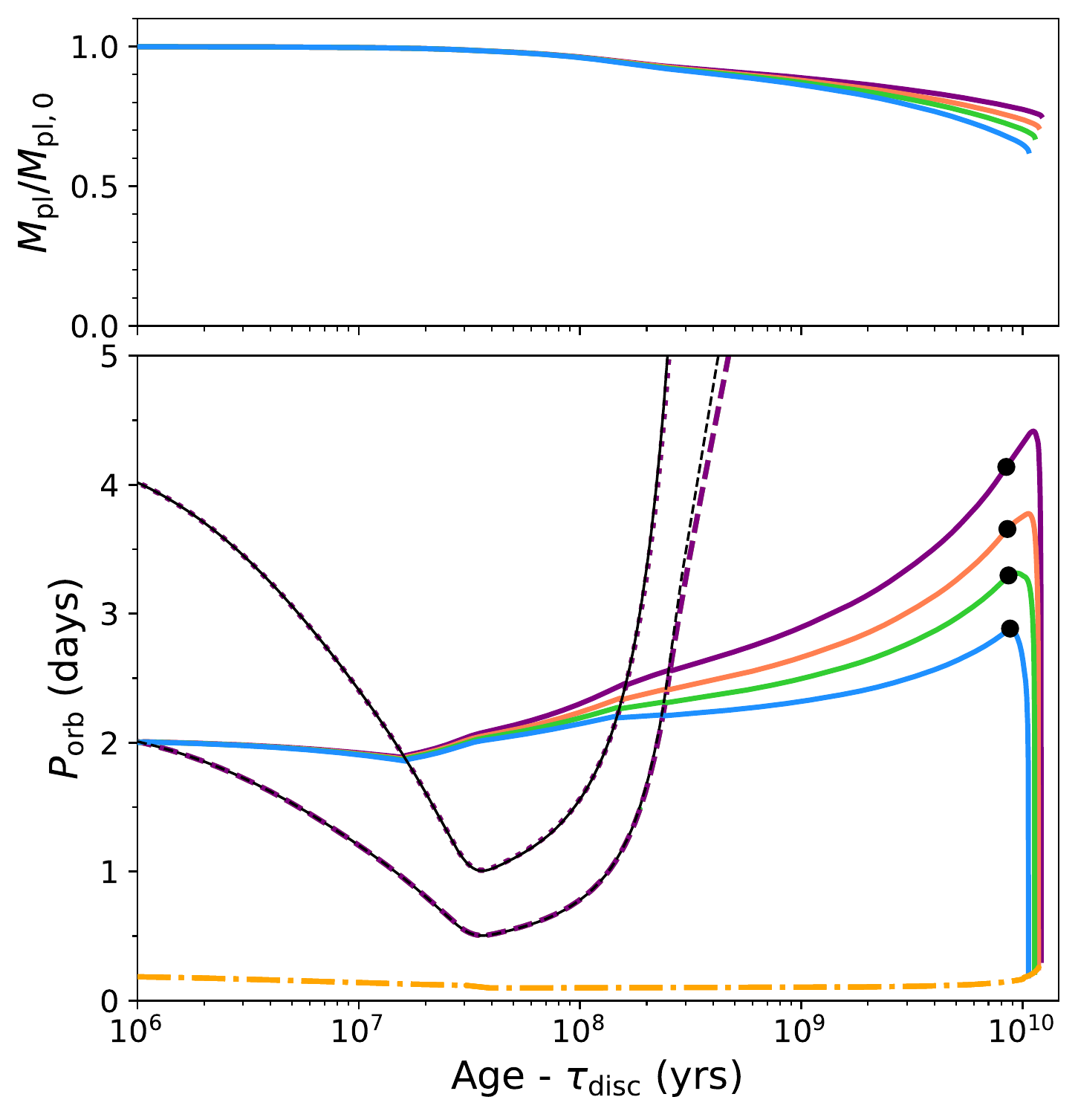}\par 
	\includegraphics[width=\columnwidth]{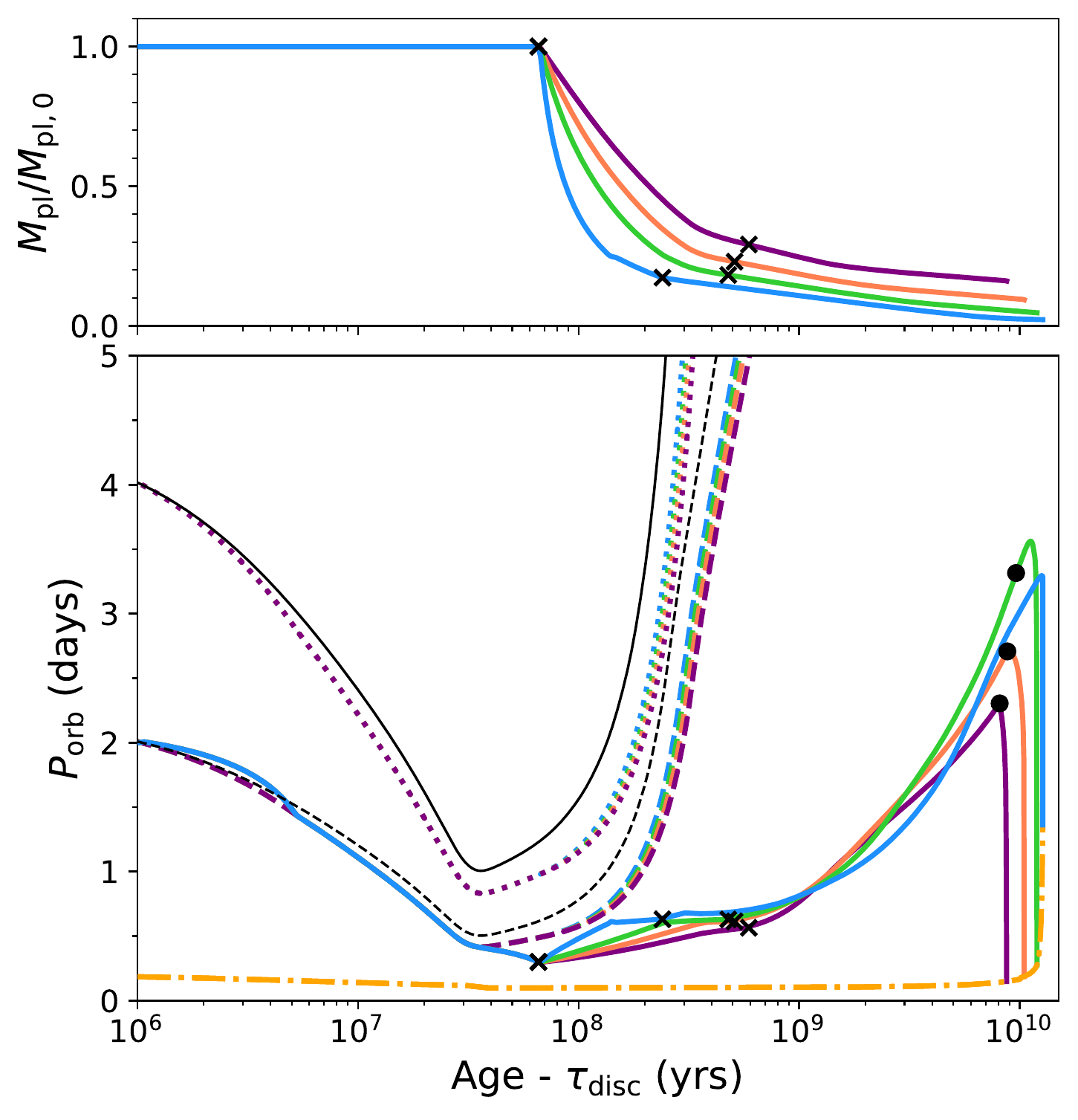}\par
    \end{multicols}
        \caption{Same as Fig.~\ref{fig3}, but for $M_\mathrm{pl, 0} = 0.3 \;M_\mathrm{J}$, $P_\mathrm{rot,0}$ = 4.5 days (left panel) and $M_\mathrm{pl, 0} = 2\;M_\mathrm{J}$, $P_\mathrm{rot,0}$ = 4.5 days (right panel). Blue, green, orange, and purple colors correspond to $\chi$ = 0.3, 0.5, 0.7, and 1, respectively.}
\label{fig7}
\end{figure*}

As we reported in L21, the initial stellar spin determines the relative contribution of inertial/gravity waves in the dynamics of hot Jupiters in the sense that inertial waves play a decisive role in shaping the distribution of planets around fast rotators, while systems with slow rotators are primarily affected by the dissipation of gravity waves. Our main conclusions about the impact of the initial rotation rate on the orbital dynamics remain partially valid within the model presented here. In Fig.~\ref{fig6}, we compare the secular evolution of systems composed of hot Jupiter with $M_\mathrm{pl} = 2\;M_\mathrm{J}$ orbiting rapidly rotating (left panel) and slowly rotating (right panel) solar-mass star. Similarly to L21, the planets around fast rotators outside the corotation radius undergo rapid outward migration, driven by the inertial wave dissipation, and remain stable against tidal inspiral until TAMS. Hot Jupiters inside the corotation radius encounter the $P_\mathrm{orb} = \frac{1}{2} P_\mathrm{rot}$ limit and stay on the edge of the inertial wave excitation region until the RLO phase begins. Intense tidal dissipation provides a short mass-loss timescale, which may lead to the violation of thermal equilibrium and planetary disruption. If the planets survive RLO despite the extremely powerful mass transfer, they subsequently cross the corotation radius and move away from a star. In contrast to the planets initially outside the corotation radius, the closest hot Jupiters merge with the host before TAMS.

The evolution around slow rotators is marginally different from our findings in L21. Implementation of the magnetic interaction makes a minor correction to the infall time, and the planets reveal negligible migration prior to the onset of gravity wave damping.

Fig.~\ref{ap1} shows the event diagrams for eight initial stellar rotation rates considered in the present research. The event diagrams (infall diagrams in L21) display the fate of a planet by TAMS, depending on its initial semi-major axis and mass. Red crosses represent the no-merger region, corresponding to the cases in which neither stable mass transfer, nor tidal disruption, nor direct impact occurs. One can see that the merger region, occupied by the rest of the Jovian planets, increases with decreasing stellar spin. Some of the most massive hot Jupiters, depicted by yellow circles, plunge directly into the stellar atmosphere before the initiation of gravity wave breaking. Lower mass planets, highlighted in purple, end up being tidally disrupted. The infalls accompanied by gravity wave dissipation, shown in green, prevail in systems with slow rotators. However, taking into account stable mass transfer significantly increases the impact of gravity waves in the orbital evolution around rapid rotators with respect to the model in L21. The RLO cases are marked by squares and triangles, depending on whether RLO begins before or after ZAMS. As demonstrated in Fig.~\ref{ap1}, the early RLO events are more common among the systems with fast rotators, which is related to the impact of inertial waves. We recall that the inertial wave dissipation efficiency is proportional to the squared angular velocity. In addition, given that, with increasing stellar spin, the bottom limit of the inertial wave excitation (the  $P_\mathrm{orb} = \frac{1}{2} P_\mathrm{rot}$ limit) gets lower, the planet undergoes rapid migration at shorter orbital separations, resulting in RLO at earlier ages of stellar evolution.

During the post-RLO phase, the planetary remnant may either be completely evaporated or survive until TAMS or merge due to gravity wave breaking. These scenarios are represented by blue, black, and green colors, respectively. If mass transfer before ZAMS is unstable, one might expect the lack of hot Jupiters orbiting rapid rotators, which is confirmed by the observations (\citealt{McQuillan}), although the latter feature may be attributed to a scarcity of data (\citealt{Messias}).

\begin{figure*}
\begin{multicols}{2}
	\includegraphics[width=\linewidth]{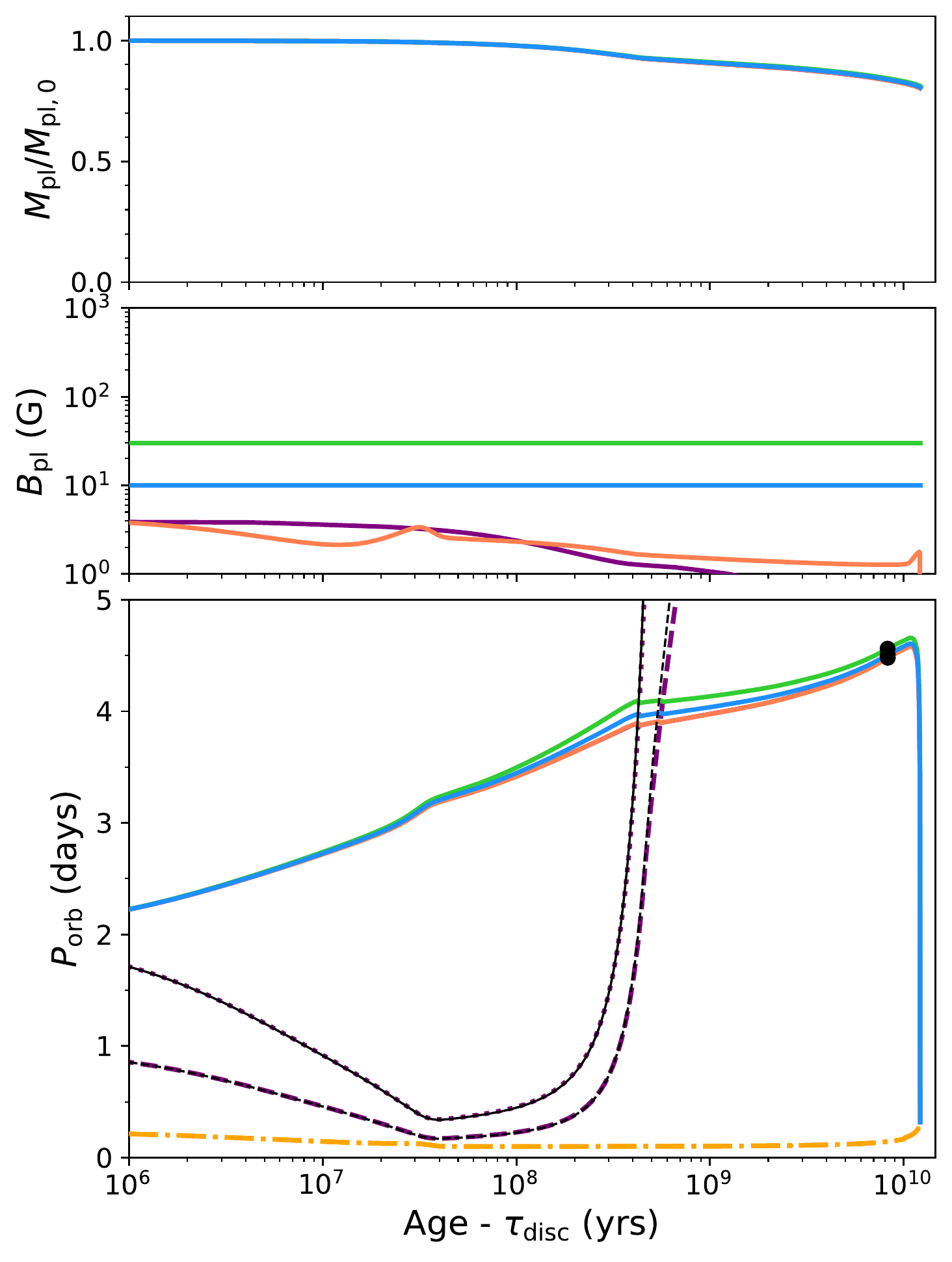}\par 
	\includegraphics[width=\columnwidth]{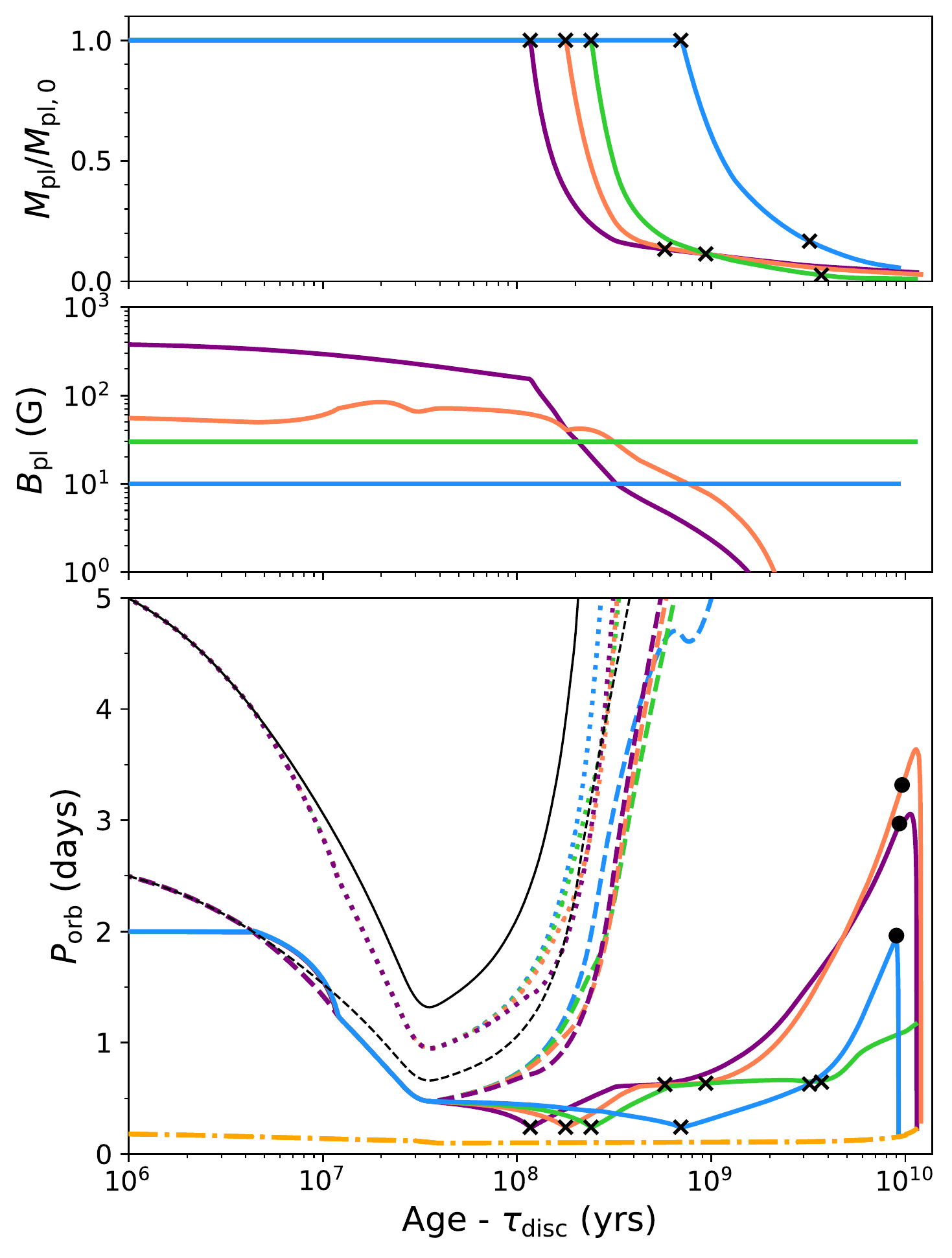}\par 
    \end{multicols}
        \caption{Same as Fig.~\ref{fig3}, but for $M_\mathrm{pl, 0} = 0.3 \;M_\mathrm{J}$, $P_\mathrm{rot,0}$ = 2 days (left panel) and $M_\mathrm{pl, 0} = 3\;M_\mathrm{J}$, $P_\mathrm{rot,0}$ = 5.5 days (right panel). Middle panel: the equatorial magnetic field strength at the planetary surface $B_\mathrm{pl}$. Blue and green colors correspond to the constant $B_\mathrm{pl}$ = 10 and 30 G, respectively. Orange color represents the incident flux-dependent model by~\protect\cite{Yadav}, and purple color represents the internal flux-dependent model by~\protect\cite{Hori}.}
\label{fig8}
\end{figure*}

\subsection{Impact of the angular momentum loss}
\label{subsec:chi}
As discussed in subsection~\ref{subsec:phot}, the orbital expansion driven by the mass loss is sensitive to the fraction of the orbital angular momentum conserved within the orbit, denoted by the parameter $\chi$.  For the conservative mass transfer, the parameter $\chi$ equals unity. Note that the high fraction of the angular momentum returned to the orbit implies its strong expansion to compensate for the mass loss. Moreover, the value of $\chi$ defines the condition for stability of RLO, namely, the requirement of a positive denominator in eq.(\ref{eq:RLO2}). Small $\chi$ results in a weak orbital response to the mass transfer. If the planetary radius increases with mass loss, the planet may fall below the Roche limit, eventually leading to dynamically unstable RLO and planetary disruption. Below we test our migration model with $\chi$-values of 0.3, 0.5, 0.7, and 1, which satisfy the stability criterion. 

The left panel of Fig.~\ref{fig7} visualizes the effect of changing $\chi$ on the low-mass end of the hot Jupiter range. These planets lose a substantial part of the envelope via photoevaporation, and small $\chi$ hinders the increase of star-planet separation, allowing hot Jupiter to receive a higher incident flux and evaporate more intensively. Thus, the choice of $\chi$ impacts both the planetary current location and mass.

This trend continues when considering hot Jupiters undergoing RLO, demonstrated in the right panel. Again, decreasing $\chi$ causes mass loss enhancement. The planet with $\chi$ = 0.3, shown in blue, loses most of its envelope, which predetermines the absence of gravity wave dissipation. In contrast, the gravity wave dissipation starts to operate in systems with $\chi$ = 0.5, 0.7, and 1.0, shown in green, yellow, and purple, respectively. The former two planets remain stable until TAMS, while the latter one decays during the host MS lifetime. Therefore, we might expect $\chi$ to affect the architecture of planets suffering significant mass loss.

\subsection{Impact of planetary magnetism}
\label{subsec:magn}
Another factor influencing the secular evolution of star-planet systems is the planetary magnetic field strength $B_\mathrm{pl}$. In Fig.~\ref{fig8}, we compare the orbital tracks of hot Jupiters with different magnetic field parametrizations. Purple line corresponds to the reference prescriptions by \cite{Hori}, blue and green lines represent constant-$B_\mathrm{pl}$ model (with $B_\mathrm{pl} =$ 10 and 30 G, respectively), and orange line illustrates the model based on the prescription by \cite{Yadav}, according to which the planetary dynamo is induced by the incident flux. Left panel demonstrates the case of a planet with $M_\mathrm{pl, 0} = 0.3 \;M_\mathrm{J}$ around a rapid rotator. Before ZAMS, tidal effects dominate the magnetic interaction since high stellar spin promotes intense dissipation of the inertial waves. As the star spins down due to wind braking, the tidal forces weaken, and the magnetic field plays a more significant role as the divergence of the orbital tracks slightly increases. After crossing the corotation radius, the sign of magnetic torque changes, and the difference in star-planet separation, owing to $B_\mathrm{pl}$, reduces. All the time until TAMS, the planets are migrating outward because the mass-loss contribution prevails over the magnetic and tidal forces.

For massive hot Jupiters inside the $P_\mathrm{orb} = \frac{1}{2} P_\mathrm{rot}$ limit, magnetic interaction becomes the only essential mechanism driving inward migration, and the time at which the planet fills the Roche lobe depends primarily on the magnetic field. The right panel of Fig.~\ref{fig8} illustrates the secular evolution of hot Jupiter systems with $M_\mathrm{pl, 0} = 3.0 \;M_\mathrm{J}$ and $P_\mathrm{rot,0}$ = 5.5 days (median rotator). One can see that the prescription by \cite{Hori} gives the highest estimate of $B_\mathrm{pl}$, causing the fastest migration rates until the planet reaches the Roche limit. The magnetic field strength by \cite{Yadav} provides the second earliest RLO, while the constant-$B_\mathrm{pl}$ models lead to the latest arrival at the Roche limit. Subsequently, the magnetic field governs stable mass transfer, and its termination is a subject of the form of $B_\mathrm{pl}$-dependence on mass and age. The model with $B_\mathrm{pl} = 30$ G sustains RLO for a longer time than other prescriptions here, leaving the corresponding planet stripped off by TAMS. Other planets, shown in the right panel, remain massive enough to initiate gravity wave dissipation before the engulfment.

\begin{figure*}
\begin{multicols}{2}
    \includegraphics[width=\linewidth]{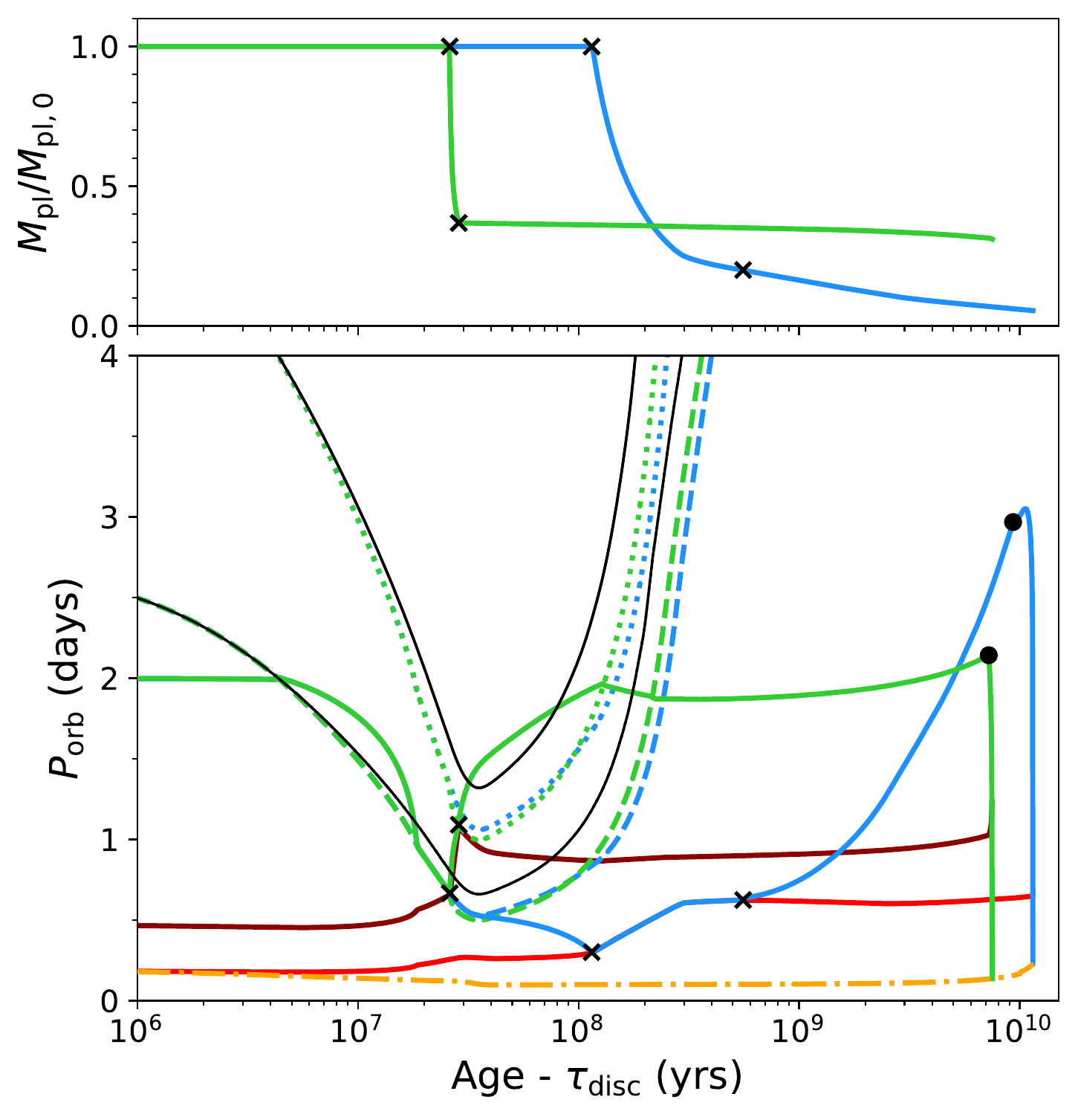}\par 
    \includegraphics[width=\linewidth]{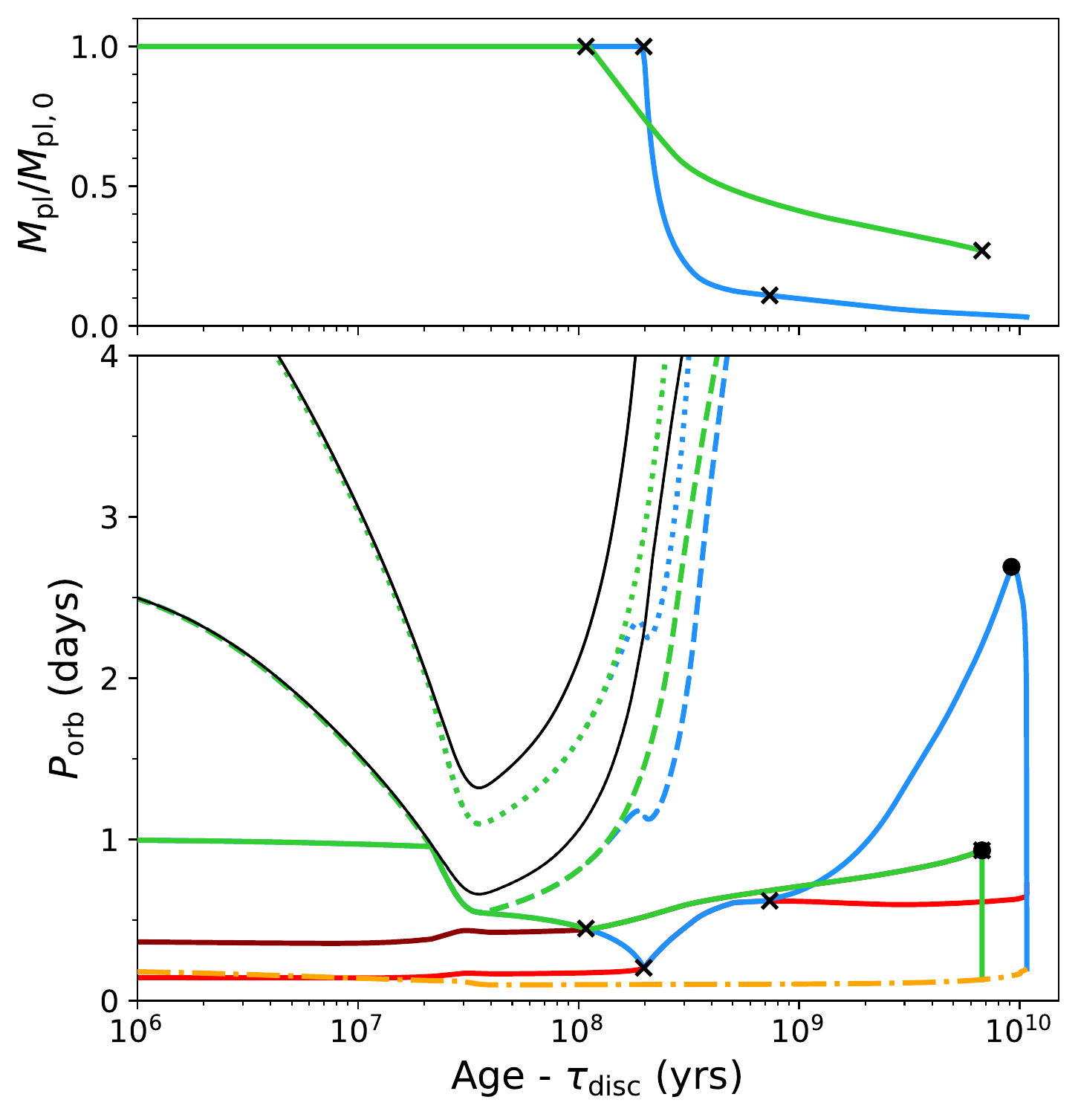}\par  
    \end{multicols}
\caption{Same as Fig.~\ref{fig3}, but for $M_\mathrm{pl, 0} = 2 \;M_\mathrm{J}$, $P_\mathrm{rot,0}$ = 5.5 days (left panel) and $M_\mathrm{pl, 0} = 4\;M_\mathrm{J}$, $P_\mathrm{rot,0}$ = 5.5 days (right panel). Dark red line is the refined Roche limit, and green lines illustrate the associated case of secular evolution of a star-planet system.}
\label{fig10}
\end{figure*}

\subsection{Impact of the Roche limit definition}
\label{subsec:rlim}
The Roche limit, $a_\mathrm{R}$, is another source of uncertainty regarding overflow simulation. This uncertainty reveals itself through the parameter $f_\mathrm{p}$ defining $a_\mathrm{R}$ via the eq. (\ref{eq:rlimit}). Throughout the previous subsections we set $f_\mathrm{p} = 3^{1/3}$. Effectively, this is equivalent to approximating the Roche lobe with the Hill sphere, which is justified for a perfectly spherical planet. Nonetheless, RLO may begin earlier. For example, from three-dimensional hydrodynamical simulations, \cite{Guillochon} inferred that Jupiter-like planets are disrupted by tidal forces at $f_\mathrm{p} = 2.7$. We adopt the following value to test how our model reacts to the corresponding Roche limit transformation. 

With the extended $a_{R}$, the planets initiate mass transfer earlier. This feature is demonstrated in both panels of Fig.~\ref{fig10}, where the green lines represent the refined model, while the blue lines correspond to the reference prescription. As a consequence of the applied changes, more hot Jupiters tend to fill the Roche lobe before ZAMS when stellar density is lower. The latter is reflected in a higher amount of tidal disruption and direct impact events as the condition $\rho_\mathrm{pl} < \rho_{*}$ required for stable mass transfer becomes harder to satisfy.

The second crucial difference concerns the intensity of RLO. If overflow occurs at a higher star-planet separation, the mass exchange is expected to be less effective since tidal and magnetic forces responsible for the removal of the planetary orbital angular momentum weaken with distance from the host. As a result, the mass-loss timescale may become comparable with stellar MS lifetime, the case depicted in the right panel. The further the Roche limit is, the more massive the planet remains once gravity waves begin to dissipate. In turn, the low value of $a_\mathrm{R}$ increases the probability of a gas giant's transformation into a hot Neptune or super-Earth at some point in its dynamical evolution.

\begin{figure*}
\begin{multicols}{2}
    \includegraphics[width=\linewidth,height=5.4cm]{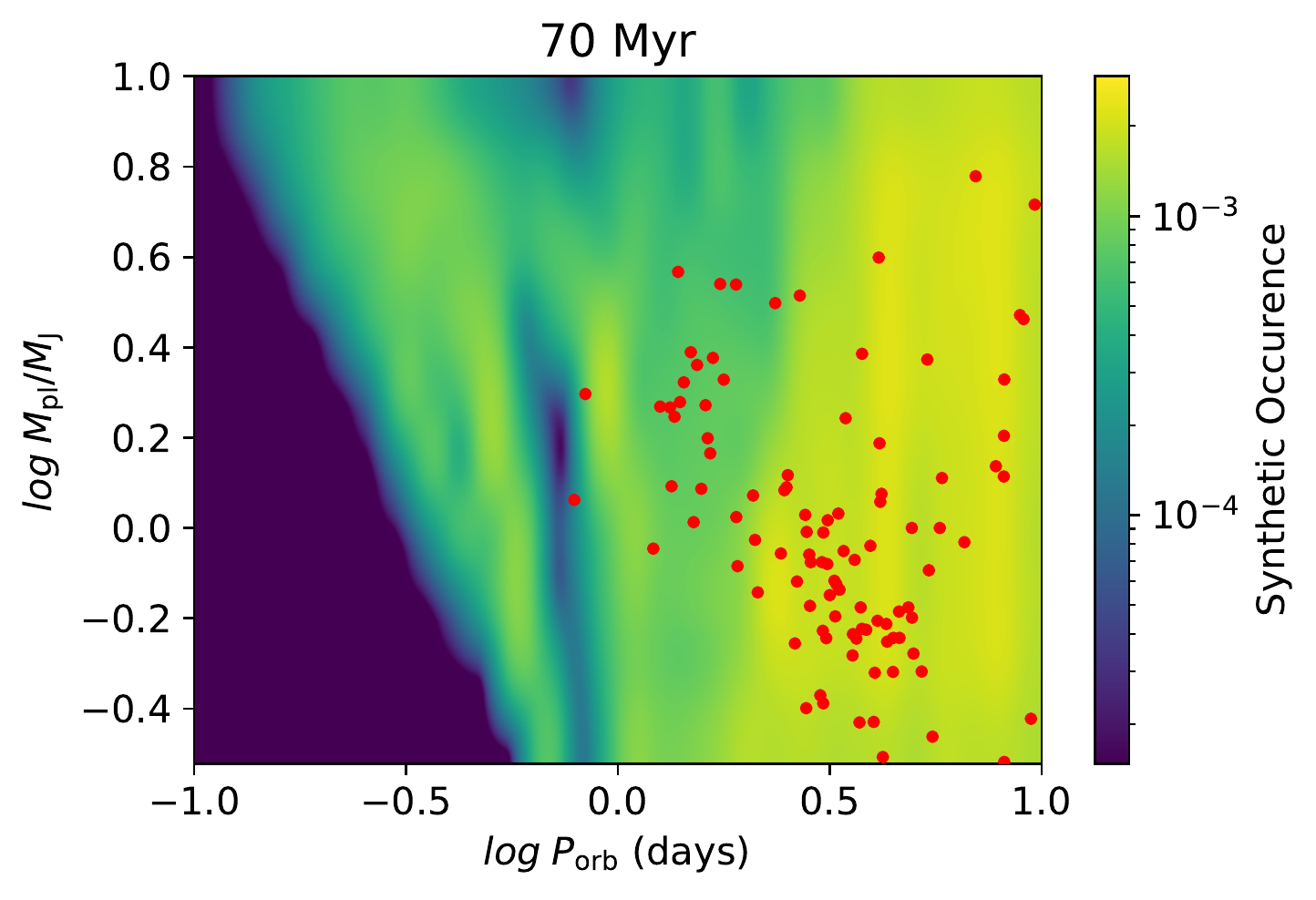}\par 
    \includegraphics[width=\linewidth,height=5.4cm]{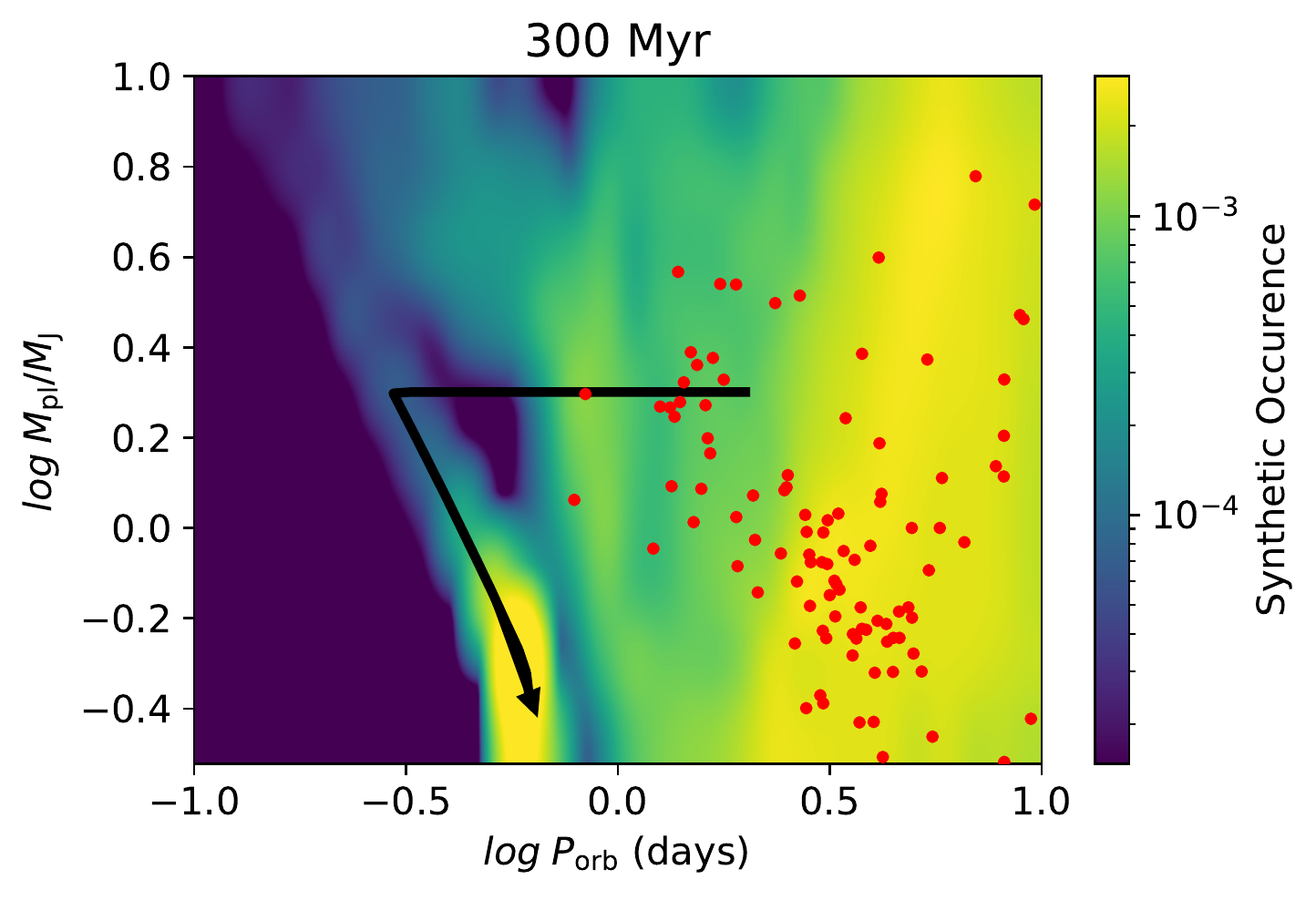}\par
    \includegraphics[width=\linewidth,height=5.4cm]{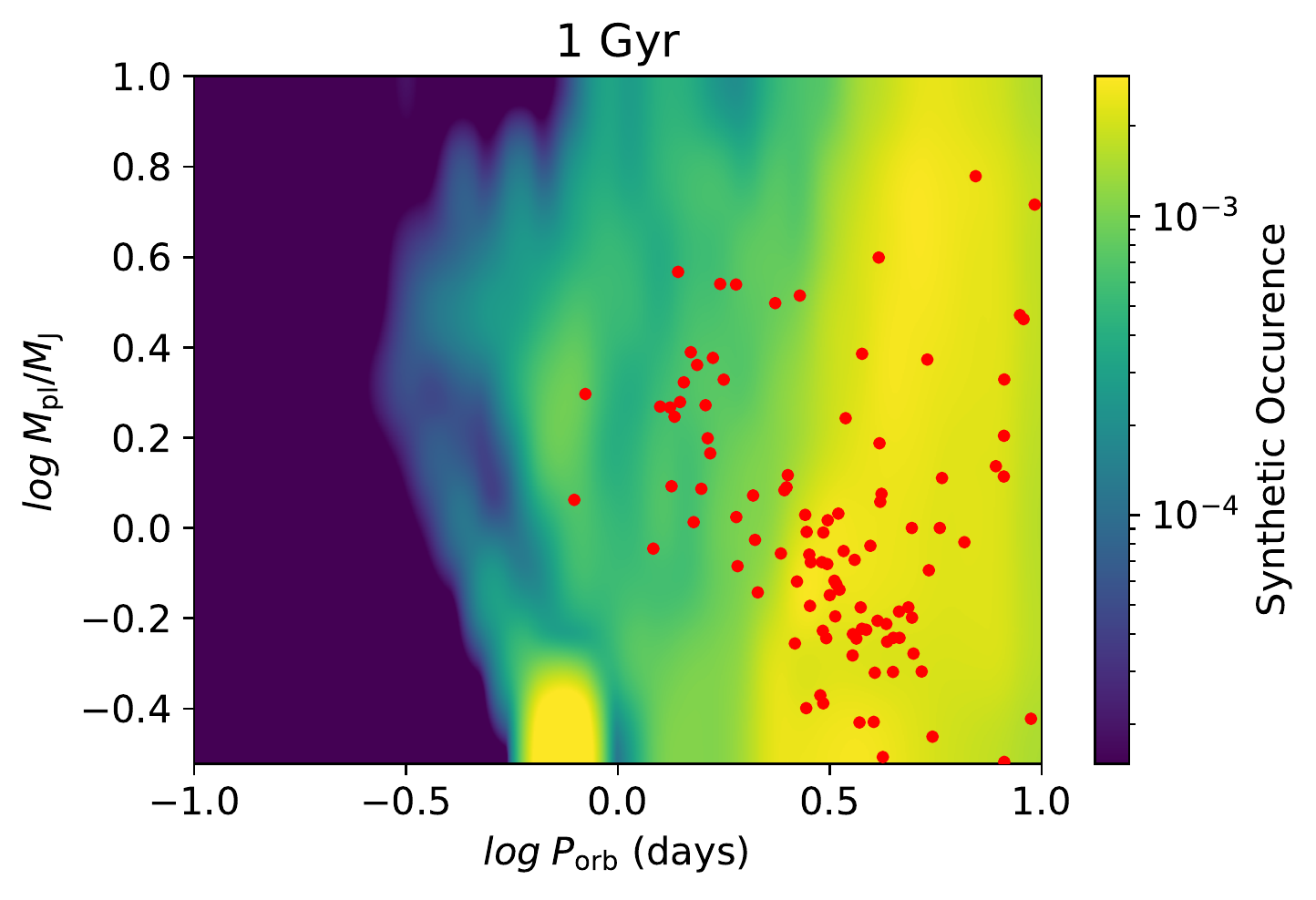}\par 
    \includegraphics[width=\linewidth,height=5.4cm]{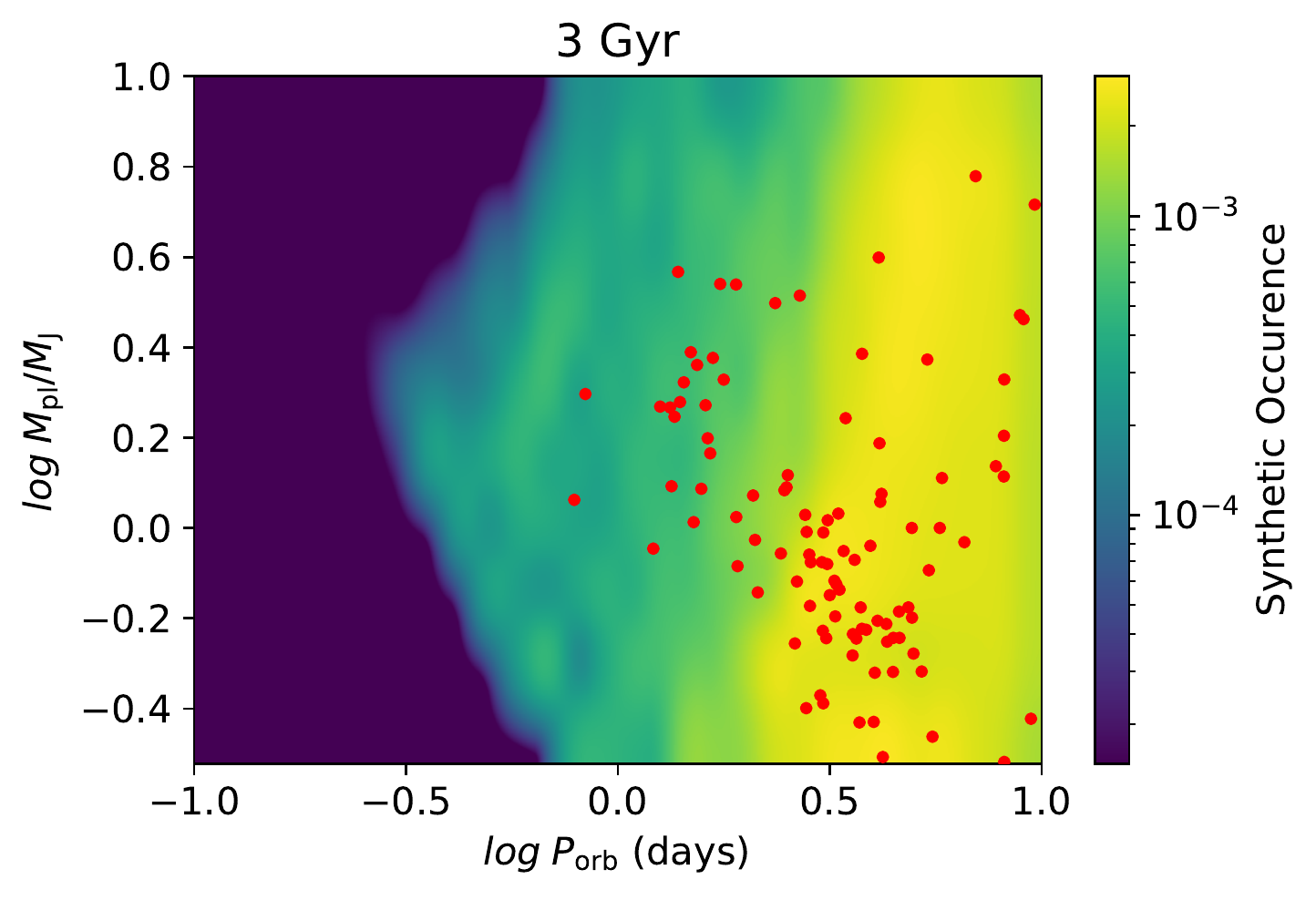}\par
    \includegraphics[width=\linewidth,height=5.4cm]{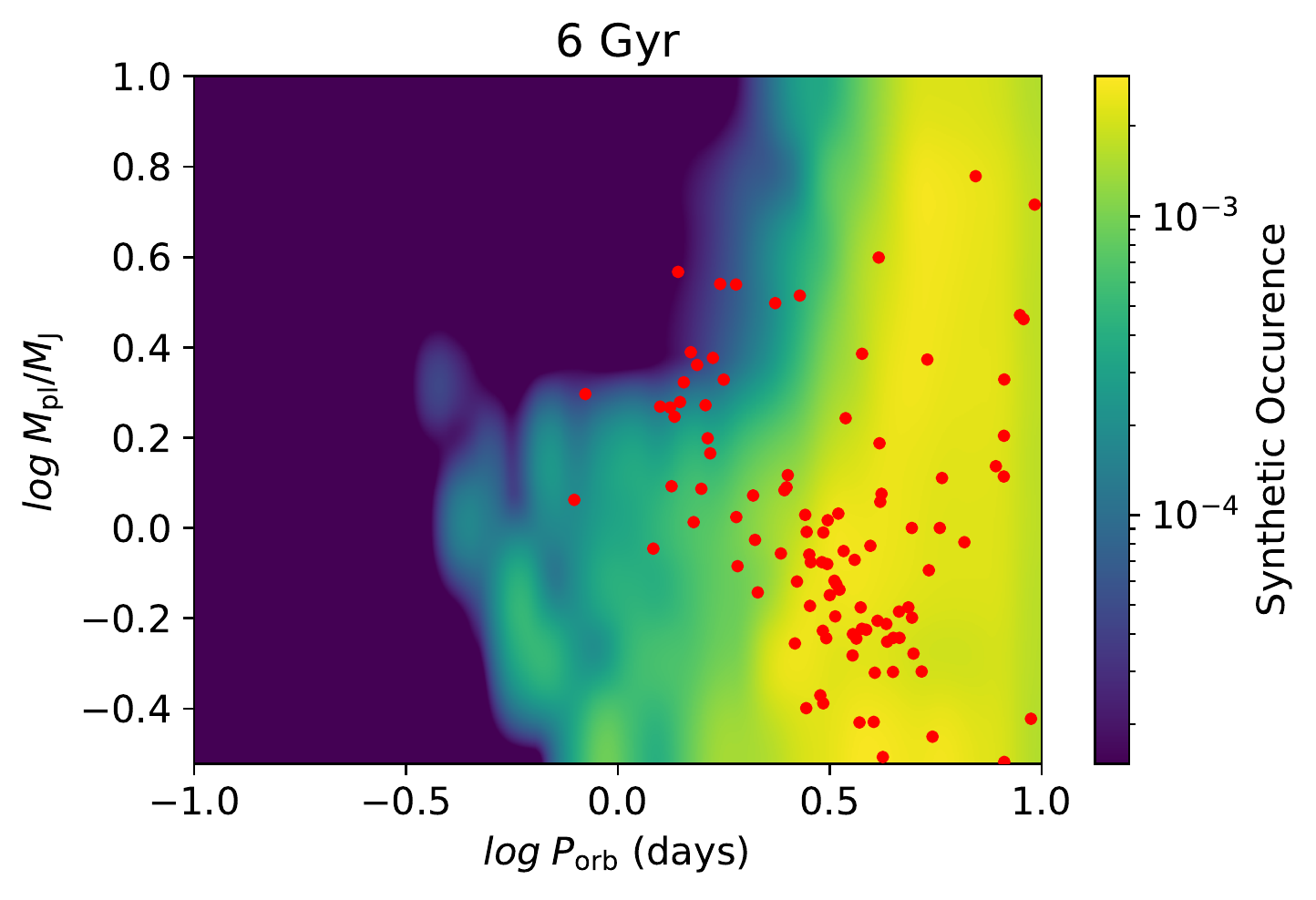}\par 
    \includegraphics[width=\linewidth,height=5.4cm]{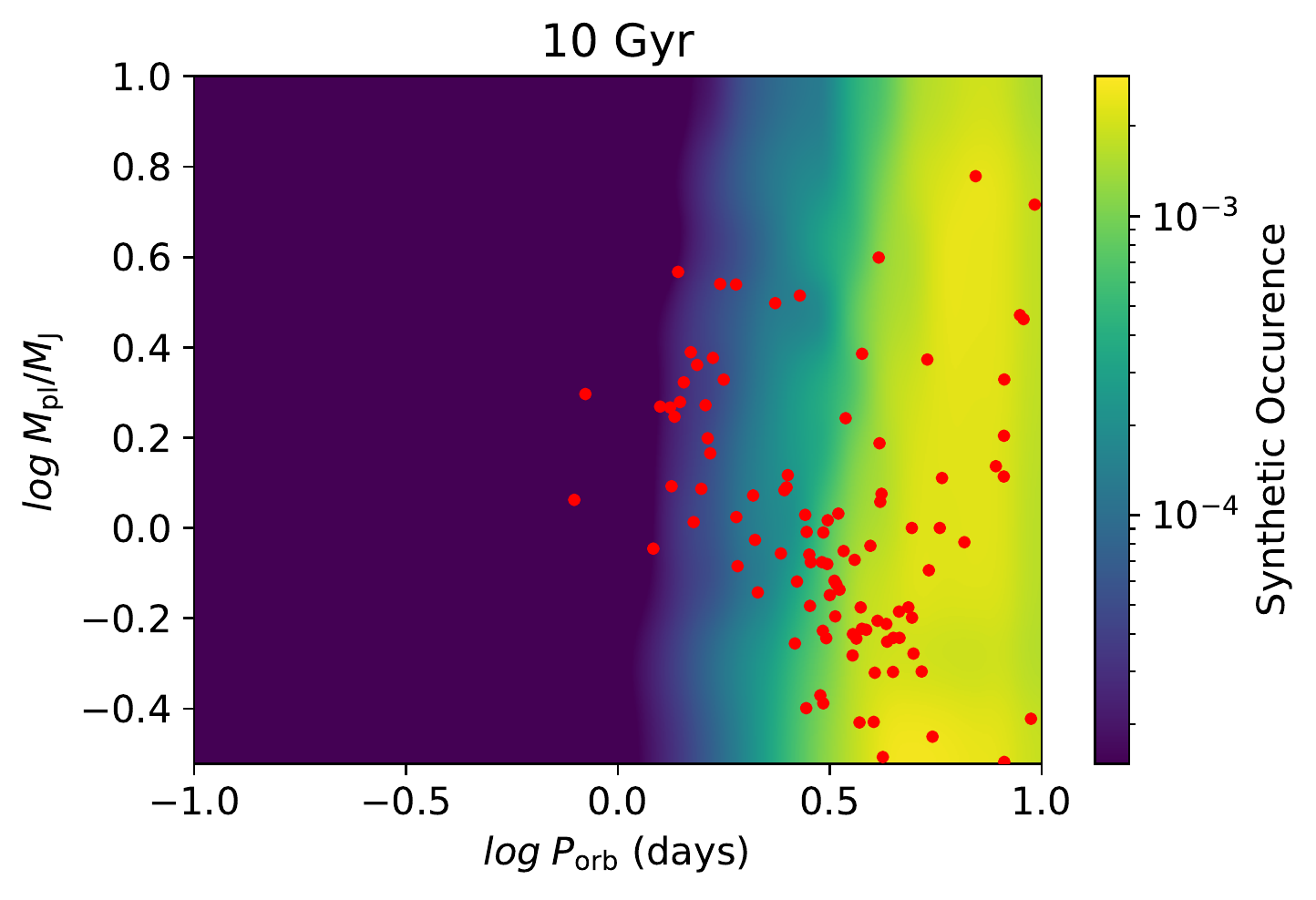}\par
    \end{multicols}
    \vspace*{-6mm}
    \caption{Occurrence rate density of simple hot Jupiter population at different ages. Black arrow is the trajectory of the reference star-planet system, shown in Fig.~\ref{fig1}. Red circles: observational sample of transit planets around solar-mass stars from the NASA Exoplanet Archive (\url{https://exoplanetarchive.ipac.caltech.edu/}).}
\label{fig11}
\end{figure*}

\section{Application to the population synthesis}
\label{sec:popsynth}
We now aim to understand how the processes implemented in the present study affect the hot Jupiter population. To do so, we begin with simple constant-age model based on the log-uniform distributions of initial planetary mass and semi-major axis and then dwell on a more sophisticated model taking into account the detection probability and star formation history. Using the grid of pre-computed star-planet simulations, we determine the fate of a system with arbitrary parameters at a given age. Every synthetic population is composed of $10^6$ planets. Contrary to L21, we explore the statistical patterns associated with hot Jupiters around solar-mass stars only. Throughout the following section, the distribution of the initial stellar rotation period $P_\mathrm{rot,0}$ is adopted from L21:
\begin{equation}
  p(\log \; P_\mathrm{rot,0})    \propto \exp\left(-\frac{(\log \;P_\mathrm{rot,0} - \zeta_*)^2}{2\sigma_*^2}\right),
    \label{eq:pdistribution}
\end{equation}
with $\zeta_* = 0.81$, $\sigma_* = 0.24$. Although the above expression is obtained using the different disc lifetime parametrization, applying eq.(\ref{eq:disc}) changes the final law within the margin of error. When simulating the initial rotation period, we exclude stars with $P_\mathrm{rot,0} < 2$ days.

\subsection{Simple model of hot Jupiter population}
\label{subsec:model0}
In the first step, we consider a simple constant-age model of the initial hot Jupiter population. The orbital period distribution is log-uniform in the range between 1 and 10 days, while planetary mass is drawn from a log-uniform distribution between 0.3 and 10 $M_\mathrm{J}$. We assume that every hot Jupiter is observed regardless of its current location. Although the corresponding population is far from the observed one, it provides an opportunity to explore the key properties underlying the evolution of hot Jupiter systems.

Snapshots of a simple population at different ages are shown in Fig.~\ref{fig11}. The occurrence rate density here is normalized to the fraction of the remaining hot Jupiters with respect to the initial population. We intend to find how well our simple model reproduces the observed distribution. To do this, we select from the NASA Exoplanet Archive all transit close systems composed of hot Jupiter and a solar-mass star ($\mathrm{0.95\;M_\mathrm{\odot} < M_\mathrm{*} < 1.05\;M_\mathrm{\odot}}$). The corresponding sample is shown with red circles.  

In the first snapshot, displaying the population shortly after ZAMS, the synthetic distribution is already extended toward low orbital periods as massive planets around fast and median rotators are delivered close to the host due to inertial wave dissipation. These hot Jupiters subsequently undergo RLO, as demonstrated in the middle left panel. The arrow illustrates the first reference case described in subsection~\ref{subsec:ref}. While transferring mass, hot Jupiter moves along the tail, representing the Roche limit. In our model, the stable mass transfer usually ends before 1 Gyr, with the planets suffering orbital expansion driven by photoevaporation. One can see a cluster of Roche-lobe overflowing objects and post-RLO remnants a little further from the Roche limit, visible in the middle and bottom left panels. Thermally driven outflow moves these planets to the lower right side of the diagram so that the cluster dissolves by 3 Gyr, roughly the mean age of the Galactic thin disc population (\citealt{Haywood}). Note that the occurrence of hot Jupiters in the region of the low orbital period is reduced compared with the initial log-uniform distribution. This change, however, is not enough to match the present-day observational sample. Soon after 3 Gyr, gravity waves begin to dissipate in stars hosting the most massive planets considered in the present paper. The resulting planetary infall develops a boundary in the upper part of the diagram, noticeable in the middle right panel. The infalls driven by gravity wave breaking can potentially explain the observed scarcity of low-period hot Jupiters with $M_\mathrm{pl} > 4\;M_\mathrm{J}$. To recreate this scarcity, one has to adjust the star formation history and initial planetary mass distribution. Nevertheless, this model cannot reproduce the lower edge of Jovian planets since tidal and magnetic interactions, as well as photoevaporation, are insufficient to clear out the respective region of the diagram, in agreement with \cite{Vissapragada}. Only after 10 Gyr, shortly before TAMS, the low-mass part of the synthetic population is shaped in a way that partially resembles the edge of the present-day distribution.
\subsection{Detailed model of hot Jupiter population}
\subsubsection{Model 1}
\label{subsec:model1}
To make the quantitative predictions regarding planetary infall, the simple model needs to be improved. First, we adopt the uniform stellar age distribution between 0 and 7 Gyr to reproduce the Galactic thin disc population in accordance with star formation rate (SFR) history from \cite{Haywood}. Same as in L21, we neglect the thick disc population since the observed anti-correlation of hot Jupiter occurrence rate with stellar metallicity (\citealt{Petigura}) and low average abundance of thick disc population (\citealt{Gilmore}) implies that its contribution is marginal. 

Second, in the new population, the planets producing full transit are highlighted. The geometric transit probability is $p_\mathrm{transit} = 0.7R_*/a$, as in \cite{Fulton}. Factor 0.7 is introduced to exclude grazing planets.

We demonstrated in \ref{subsec:model0} that our migration model is not able to reproduce the observed lower edge of sub-Jovian planets in the mass--separation diagram. Following \cite{Bailey}, we suggest that this boundary may be associated with the protoplanetary disc's inner edge, defined by the truncation radius $R_\mathrm{t}$: 

\begin{equation}
R_\mathrm{t} = \left(\frac{\mathcal{M}^2\tau_\mathrm{ac}}{M_\mathrm{pl}\sqrt{GM_{*}}} \right)^{2/7},
    \label{eq:trunk}
\end{equation}
where $\tau_\mathrm{ac} \sim 10^{5}$ yrs is characteristic accretion timescale and $\mathcal{M} = B_* R_*^3$ is the stellar magnetic moment at T-Tauri phase. The values of $B_*$ and $R_*$ are adopted from \cite{Bailey} (1 kG and 1.2 $R_{\odot}$, respectively). The initial orbital period is drawn from a log-uniform distribution between max(1.0, $P_\mathrm{t}$) and 10 days, with $P_\mathrm{t} = 2\pi \sqrt{\frac{R_\mathrm{t}^3}{GM_{*}}}$.

To simulate the initial planetary mass, we address the distribution from L21 based on the hot Jupiter sample around FGK-stars from the NASA Exoplanet Archive:
\begin{equation}
   p\left(\log \frac{M_\mathrm{pl}}{M_\mathrm{J}}\right) \propto \begin{cases}
   \exp\left(-\frac{(\log \frac{M_\mathrm{pl}}{M_\mathrm{J}} - \zeta_\mathrm{pl})^2}{2\sigma_\mathrm{pl}^2}\right), & \log\; \frac{M_\mathrm{pl}}{M_\mathrm{J}} \leq 0.3\\
    \left(\log \frac{M_\mathrm{pl}}{M_\mathrm{J}}\right)^{\beta_\mathrm{pl}}, & \log \frac{M_\mathrm{pl}}{M_\mathrm{J}} > 0.3
 \end{cases},
    \label{eq:pl_mass}
\end{equation}
where $\zeta_\mathrm{pl} = -0.067$, $\sigma_\mathrm{pl} = 0.31$, and $\beta_\mathrm{pl} = -1.46$. Note that the above fit matches the present-day observational data, however, we intend to obtain the initial distribution. In order to achieve this, we recalculated the coefficients  $\zeta_\mathrm{pl}$, $\sigma_\mathrm{pl}$, and $\beta_\mathrm{pl}$ to make the simulated present-epoch transit population conform to the sample of known hot Jupiters. Thus we obtain $\zeta_\mathrm{pl} = -0.044$, $\sigma_\mathrm{pl} = 0.29$, and $\beta_\mathrm{pl} = -1.02$. The following coefficients establish the planetary mass distribution for Model 1.

\subsubsection{Model 2}
\label{subsec:model2}
Our second model relies on the same considerations regarding the systems' age and observability. However, we now move from reference to the modified Roche limit from subsection~\ref{subsec:rlim}. We assume that the initial distribution of the orbital period is uniform in the logarithm between max(1.0, $P_\mathrm{R}$) and 10 days, where $P_\mathrm{R} = 2\pi \sqrt{\frac{a_\mathrm{R}^3}{GM_{*}}}$. This approach results in the following coefficients for planetary mass distribution: $\zeta_\mathrm{pl} = -0.018$, $\sigma_\mathrm{pl} = 0.24$, and $\beta_\mathrm{pl} = -0.74$. We refer to the corresponding model as Model 2.

\begin{figure*}
\begin{multicols}{2}
    \includegraphics[width=\linewidth,height=8.0cm]{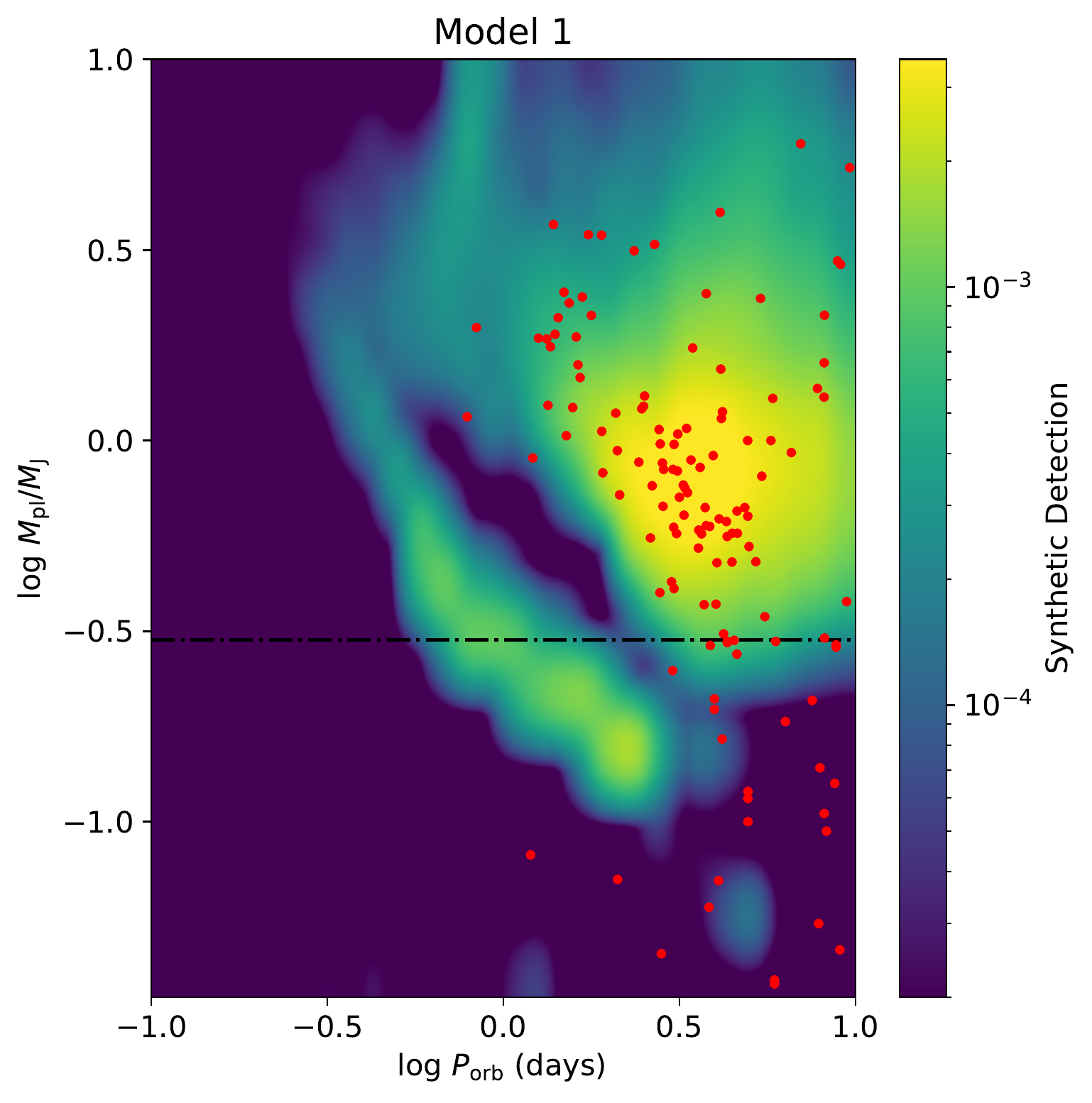}\par 
    \includegraphics[width=\linewidth,height=8.0cm]{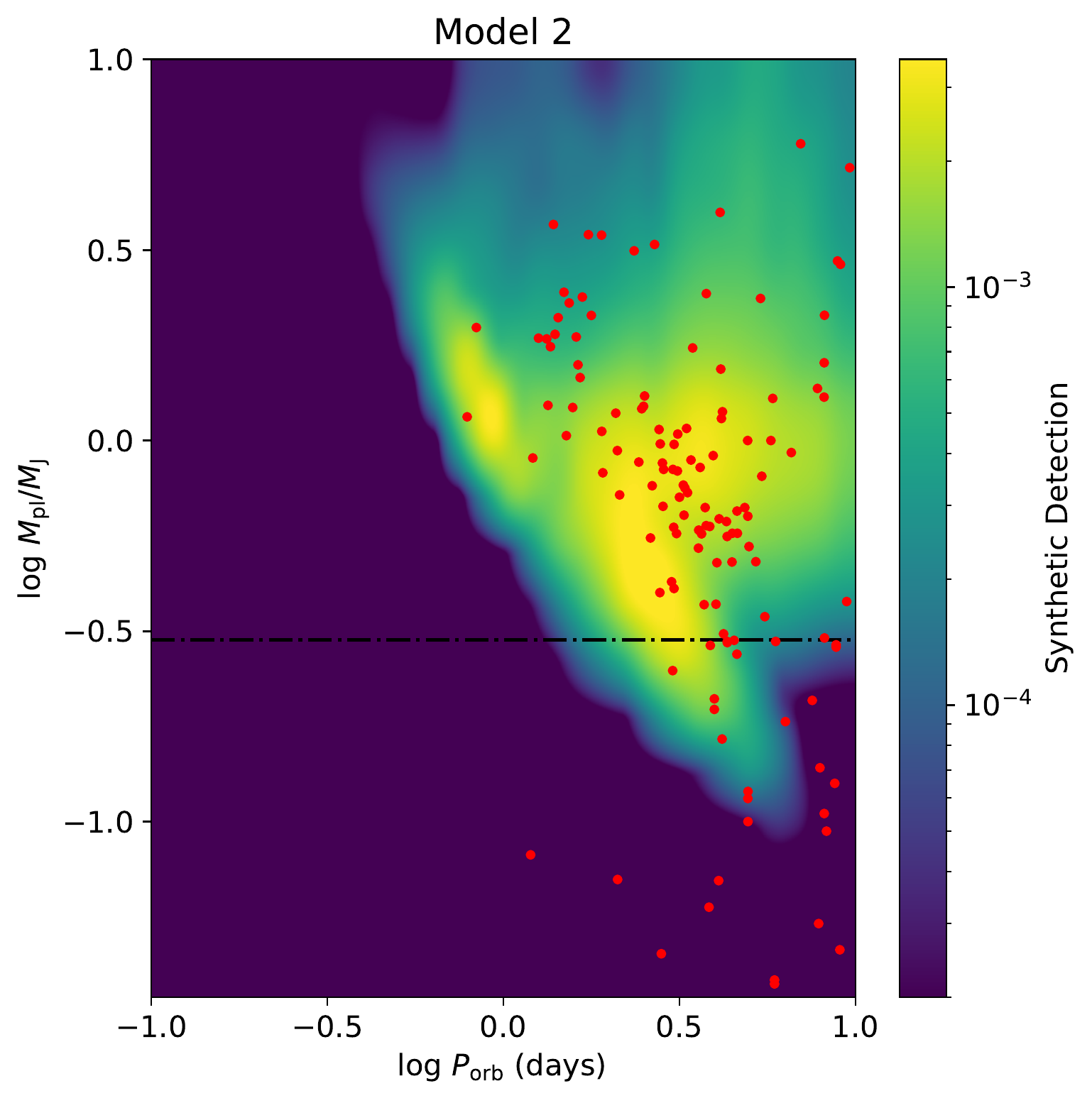}\par
    \end{multicols}
    \caption{Synthetic detection of two advanced models of hot Jupiter population. Left panel represents the model where the initial population is shaped by the inner disc radius (Model 1). Right panel corresponds to the refined parametrization of the Roche limit (Model 2). Red circles: observational sample of planets around solar-mass stars. Dash-dotted line: the lower limit of Jovian planet mass range.}
\label{fig12}
\end{figure*}

\subsubsection{Results}
\label{subsec:results}
In Fig.~\ref{fig12}, synthetic samples of transit hot Jupiters are displayed above the dash-dotted line, which indicates the lower boundary of Jovian planet mass range. Both models show reasonably good agreement with the observational data. The pile-up of hot Jupiters around $P_\mathrm{orb} \sim 3$ days is visible in both panels. We did not intend to recover the right boundary of hot Jupiters as it is likely to originate from the effects related to the planetary formation or the migration inside the protoplanetary disc. Regarding the upper edge of the synthetic distribution in the $P_\mathrm{orb}$--$M_\mathrm{pl}$ space, the scarcity of data challenges drawing definite conclusions. We speculated in subsection~\ref{subsec:model0} that, after 3 Gyr, the dissipation of gravity waves starts to clear the corresponding part of the diagram, implying that, to make the boundary better pronounced, one needs to adjust the age of the simulated systems making the population more dynamically evolved.

The main difference between the two models is the location of Jovian planets undergoing or having undergone RLO. In Model 1, these planets form the tail separated from the main cluster of planets. However, none of the 101 observed hot Jupiters around solar-mass stars, represented by red circles, occupy this tail, suggesting they are not currently filling their Roche lobe (although WASP-19b is very close to achieving this state). Given that 3.6\% of synthetic transit hot Jupiters are in the tail, the probability of not detecting an object there is relatively small, 2.4\%. One way to resolve this discrepancy is to modify SFR history to increase the mean age of the population. Alternatively, we can assume that the mass transfer becomes unstable below some critical planetary mass value. According to \cite{Jia}, the planets below 1.6 $M_\mathrm{jup}$ lose too much angular momentum to avoid disruption on a short timescale. Even if we set this critical value equal to 0.5 $M_\mathrm{jup}$, the probability of not finding a system during the RLO phase increases by more than an order of magnitude, up to 23.6\%.

At the same time, the issue concerning hot Jupiters following the Roche limit disappears when moving to Model 2, where the conglomeration of Roche-lobe overflowing objects is embedded in the main planetary cluster. The lower boundary of the population in the mass--separation diagram is qualitatively reproduced. The fraction of transit hot Jupiters undergoing RLO is 6.7\%, which is over three times higher compared with Model 1. Three of the closest observed hot Jupiters in our sample, WASP-19b, HATS-18b, and OGLE-TR-56-b, are likely to be in the process of stable mass transfer. Another two planets, namely WASP-4b and COROT-1b, might be close to filling their Roche lobes.

Below the dash-dotted line, we illustrate the planets that have abandoned the Jovian planet mass range. Given that the initial distribution of hot Neptunes and super-Earths is not simulated in this research, the corresponding part of the diagram represents only some fraction of the observed sample. At the current state of our knowledge, we cannot be confident about the relative contribution of hot Jupiter remnants to the overall population of lower-mass planets as the mechanisms underlying their formation and the early migration inside the protoplanetary disc are still uncertain. \cite{OwenLai} proposed hot Jupiters and lower-mass close-in planets having distinctly separate formation channels, which makes the direct comparison even more problematic. Nevertheless, one can see that Model 2 can potentially explain the origin of several observed hot Neptunes with $M_\mathrm{pl} > 0.1\;M_\mathrm{jup}$. In Model 2, the transformation of hot Jupiters into planets with $M_\mathrm{pl} < 0.1\;M_\mathrm{jup}$ takes longer than the adopted thin disc lifetime, meaning that the super-Earths cannot be post-RLO remnants. Though, in Model 1, the mass-loss rates are higher as the Roche limit is closer to the stellar surface, the predicted pathway to super-Earths is not occupied by any of the observed planets, which is difficult to explain if we suggest that some super-Earths are initial hot Jupiters.

According to Model 1 (Model 2), 12\% (15\%) of the initial population merged with the host or left the Jovian planet mass range before the present day. This fraction is a few times higher than the value obtained in L21, which can be easily understood. In the case of Model 1, the implementation of the magnetic interaction allows for the delivery of more close-in planets to the Roche limit prior to the beginning of gravity wave damping. In the L21 model, most of these hot Jupiters would have avoided RLO because equilibrium tide dissipation does not induce significant migration. Regarding Model 2, a big difference is made by a higher location of the Roche limit itself, making the initiation of RLO before ZAMS common.

We also calculated the fraction of transit hot Jupiters undergoing decay intense enough to be detected. Similarly to L21, we selected the systems for which the cumulative shift in transit times exceeds 5 seconds over a 10-year baseline, which corresponds to the best transit-timing observations following a single observing season (\citealt{Collier}). The cumulative shift $T_\mathrm{shift}$ in transit times after $T$ years is derived using the equation from \cite{Birkby}:
\begin{equation}
T_\mathrm{shift} = -\frac{\dot{P}}{2P_\mathrm{orb}} T^2.
    \label{eq:decay}
\end{equation}

In contrast to L21, decays proceeding before the onset of gravity wave breaking have also been included. In Model 1 (Model 2), the probability of detecting a transit time shift while observing a single hot Jupiter system for a decade is 0.25\% (0.23\%). Both estimates are in agreement with the absence of the confirmed decaying hot Jupiters orbiting a solar-mass star (the only decaying planet to date, WASP-12b, is likely to orbit a $\sim 1.4\;M_{\odot}$ subgiant star, see \cite{Weinberg1} and \cite{BaileyGoodman}). Nevertheless, the growth of the number of detected Jovian planets and the accumulation of observations are likely to result in the discovery of a decaying system in the forthcoming years.

\section{Summary}
\label{sec:summary}
In this paper, we have presented the calculations of the orbital evolution of hot Jupiters under the effect of tidal and magnetic interactions and mass-loss through photoevaporation and Roche-lobe overflow (RLO). To some extent, our migration model is the combination of approaches described in \cite{Valsecchi1}, \cite{Strugarek4}, \cite{Fujita}, and \cite{Lazovik}, allowing us to explore the secular dynamics of star-planet systems in a new fashion. In particular, our simulations predict the transformation of the Jovian planet into hot Neptune via RLO. This scenario deserves attention in the context of the objects recently found in the Neptunian desert (\citealt{West,Smith}). A similar conclusion is made in \cite{Valsecchi1} and \cite{Jackson}. However, unlike the above-mentioned studies, where a planet remains bound to the Roche limit until most of the envelope is lost, our model allows for a suspension of the RLO phase once the photoevaporation prevails over the mass transfer. The different treatment of the interplay between mass-loss driven by thermally driven outflow and RLO is not the only change relative to the model by \cite{Valsecchi1}. In the present study, we adopt a more sophisticated tidal dissipation formalism, which takes into account the dissipation of equilibrium tide, inertial waves, and gravity waves, while \cite{Valsecchi1} considered a constant tidal quality factor. This improvement makes a major contribution. We note that \cite{Valsecchi1} did not study the orbital evolution during the PMS stage and assumed that the RLO phase begins at t $\sim$ 2 Gyr. In most of our simulations, hot Jupiter reaches the Roche limit shortly after ZAMS (or even before ZAMS), when inertial waves deliver a planet close to a stellar surface, or at t > 5 Gyr, when gravity wave damping starts to operate. According to the estimates from \cite{VidalBarker1,VidalBarker}, the interaction between tidal flows and convection significantly reduces effective viscosity, and, as a result, equilibrium tide dissipation is very ineffective during the MS stage. Consequently, we implement another mechanism, the magnetic interaction, which can induce a dominant torque when dynamical waves do not dissipate.

As a result of the above changes with respect to the previous research, we suggest that the range of trajectories in the mass--separation plane is much broader. Hence, some hot Neptunes observed outside their Roche limits can still be related to the initial hot Jupiter population.

Our findings indicate that the outcome of the planetary evolution is very sensitive to the initial stellar spin. Thus, the close-in planets orbiting rapid rotators are delivered to the Roche limit before ZAMS, causing violent mass loss and potentially leading to planetary disruption. In contrast, star-planet systems composed of the (initially) slowly rotating host are stable until the onset of gravity wave dissipation, and the RLO phase is likely to be late and short. Only for median rotators, the stable mass transfer events are certain and common.

The location of the Roche limit, $a_\mathrm{R}$, strongly affects the hot Jupiter population. Placing it further from the host allows more planets to initiate RLO before ZAMS. However, it provides relatively low mass-loss rates when hot Jupiter is outside the inertial wave excitation region. In the present work, we proposed two alternative $a_\mathrm{R}$ parametrizations that define the range of predictions for overflow occurrence. The scenario where $a_\mathrm{R}$ is close to the stellar surface leads to the emergence of the tail occupied by the planets undergoing stable mass transfer at $P_\mathrm{orb} < 1$ day in the left panel of Fig.~\ref{fig12}. To make this tail statistically imperceptible, one needs to increase the mean age of the population or constrain the stability of RLO. In turn, the model with extended $a_\mathrm{R}$ is characterized by the continuous planetary conglomeration. One can notice that its lower-left boundary is shifted relative to the observed distribution. Indeed, we detect very few planets with $M_\mathrm{pl} \sim 0.5\;M_\mathrm{J}$ at $P_\mathrm{orb} < 2.5$ days, while the inner boundary of our synthetic distribution based on Model 2 begins at $P_\mathrm{orb} \sim 0.5$ day.  At the same time, the tidal and magnetic interactions shape the upper part of the diagram, resulting in the paucity of close-in giant planets with $M_\mathrm{pl} > 4\;M_\mathrm{J}$.

Applying our prescriptions to the hot Jupiter population models, we obtained that 12--15 \% of gas giants were engulfed by the host or lost too much mass to stay in the mass range of hot Jupiters. Besides, 0.20--0.25\% of simulated transit hot Jupiter systems exhibit transit timing variations available for detection within 10 years of observation. The forthcoming observations look promising in this respect.

Our approach is based on the number of assumptions that need to be verified. Among them, the most important one is probably the requirement of universality and constancy of the parameter $\chi$, describing mass and specific angular momentum loss. As shown in subsection~\ref{subsec:chi}, the value of $\chi$ impacts our calculations, altering the mass and semi-major axis of a planet at a given age. Besides, it defines the stability of mass transfer during the RLO phase. In fact, we expect $\chi$ to be a function of star-planet separation, incident flux, and stellar wind strength. The value of $\chi$ characteristic of thermally driven outflow may appear lower than during RLO, although unlikely to be zero for close-in planets (\citealt{Shaikhislamov, Debrecht}). Further hydrodynamical simulations are required to clarify this issue.

In addition, we do not take into account the influence of the magnetic field on the evaporative losses. This influence is still questionable, as there is no consensus on whether planetary magnetism mitigates (\citealt{Trammell}) or enhances (\citealt{Cohen}) the outflow.

Finally, assuming circular orbits, we opted for disc migration or in-situ formation theory. We note that another hypothesis, the high-eccentricity scenario, is often invoked by various authors to reproduce the lower edge of Jovian planets in the mass--separation diagram (e.g., \cite{ValsecchiRasio1,OwenLai}). In reality, the origin of close-in gas giants may be bimodal. As discussed in \cite{DawsonJohnson}, hot Jupiters might have formed via two channels, including disc migration (or in-situ formation) and high-eccentricity migration triggered by planet-planet scattering. In this sense, the present research sheds a light on the dynamics of only a fraction of the observed population. Further studies are needed to clarify the contribution of each mechanism to the diversity of hot Jupiter systems.

\section*{Acknowledgements}

The work on the orbital evolution code is supported by the Theoretical Physics and Mathematics Advancement Foundation ``BASIS''. We acknowledge the support provided by the Ministry of Science and Higher Education of the Russian Federation grant 075-15-2020-780 (N13.1902.21.0039) in the implementation of the hot Jupiter population modeling.

We would like to thank the anonymous referee for providing the critical comments which helped us to improve the robustness and clarity of this manuscript. Special thanks to Prof. Sergei Popov for coordinating the work. We gratefully acknowledge Drs. Antoine Strugarek and Yasunori Hori for the fruitful discussion. I would also like to thank Prof. Bill Paxton and the MESA community for making this work possible. Finally, we thank Dr. Victor Réville for access to the starAML code. 

\section*{Data Availability}

The data underlying this article will be shared on reasonable request to the corresponding author



\bibliographystyle{mnras}
\bibliography{example} 








\bsp	
\label{lastpage}
\end{document}